# On detecting biospheres from chemical thermodynamic disequilibrium in planetary atmospheres

Joshua Krissansen-Totton[1], David S. Bergsman[2], David C. Catling[1]

[1]Department of Earth and Space Sciences/Astrobiology Program, University of Washington, Seattle, WA, 98195

[2]Department of Chemical Engineering, Stanford University

Contact: joshkt@uw.edu

**Abstract**

Atmospheric chemical disequilibrium has been proposed as a method for detecting extraterrestrial biospheres from exoplanet observations. Chemical disequilibrium is potentially a generalized biosignature since it makes no assumptions about particular biogenic gases or metabolisms. Here, we present the first rigorous calculations of the thermodynamic chemical disequilibrium in Solar System atmospheres, in which we quantify the available Gibbs energy: the Gibbs free energy of an observed atmosphere minus that of atmospheric gases reacted to equilibrium. The purely gas phase disequilibrium in Earth's atmosphere is mostly attributable to $O_2$ and $CH_4$. The available Gibbs energy is not unusual compared to other Solar System atmospheres and smaller than that of Mars. However, Earth's fluid envelope contains an ocean, allowing gases to react with water and requiring a multiphase calculation with aqueous species. The disequilibrium in Earth's atmosphere-ocean system (in joules per mole of atmosphere) ranges from ~20 to $2\times10^6$ times larger than the disequilibria of other atmospheres in the Solar System, where Mars is second to Earth. Only on Earth is the chemical disequilibrium energy comparable to the thermal energy per mole of atmosphere (excluding comparison to Titan with lakes, where quantification is precluded because the mean lake composition is unknown). Earth's disequilibrium is biogenic, mainly caused by the coexistence of $N_2$, $O_2$ and liquid water instead of more stable nitrate. In comparison, the $O_2$-$CH_4$ disequilibrium is minor, although kinetics requires a large $CH_4$ flux into the atmosphere. We identify abiotic processes that cause disequilibrium in the other atmospheres. Our metric requires minimal assumptions and could potentially be calculated using observations of exoplanet atmospheres. However, further work is needed to establish whether thermodynamic disequilibrium is a practical exoplanet biosignature, requiring an assessment of false positives, noisy observations, and other detection challenges. Our Matlab code and databases for these calculations are available, open source.

**Introduction**

The most interesting question about exoplanets is whether any of them host life. In recent years, significant progress has been made in the detection and characterization of the atmospheres of Jupiter and Neptune-sized exoplanets (Barman 2007; Charbonneau *et al.* 2002; Deming *et al.* 2013; Fraine *et al.* 2014; Pont *et al.* 2008; Vidal-Madjar *et al.* 2003). With the upcoming launch of NASA's James Webb Space Telescope and the construction of larger ground-based telescopes such as the European Extremely Large Telescope, it may be possible to constrain the atmospheric composition of terrestrial planets in the near future (Belu *et al.* 2011; Deming *et al.* 2009; Hedelt *et al.* 2013; Misra *et al.* 2014; Rauer *et al.* 2011; Rodler and López-Morales 2014; Snellen *et al.* 2013). Whether or not the presence of an exoplanet biosphere could be inferred remotely from these atmospheric observations needs to be carefully considered.

Life detection using remote sensing was first proposed in the 1960s and 1970s in the context of Solar System exploration (Lederberg 1965). The realization that life on Earth has profoundly influenced the geochemical environment, and in particular the composition of the atmosphere and oceans, led naturally to the suggestion that alien biospheres may be detectable remotely via their influence on atmospheric composition (Lovelock 1965; Lovelock 1975; Lovelock and Margulis 1974). More specifically, chemical disequilibrium in planetary atmospheres, such as the co-existence of two long-term incompatible species like oxygen and methane, was proposed as a possible sign of life (Hitchcock and Lovelock 1967; Lovelock 1965). It is now understood that all the bulk gases except for the inert gases in Earth's atmosphere are modulated by biology (Catling and Kasting 2007), and so it is reasonable to expect exoplanet atmospheres to be similarly perturbed away from chemical equilibrium by biogenic gas fluxes.

Chemical disequilibrium as a biosignature is appealing because unlike searching for biogenic gases specific to particular metabolisms, the chemical disequilibrium approach makes no assumptions about the underlying biochemistry. Instead, it is a generalized life-detection metric that rests only on the assumption that distinct metabolisms in a biosphere will produce waste gases that, with sufficient fluxes, will alter atmospheric composition and result in disequilibrium.

In the modern literature on exoplanets and astrobiology, atmospheric chemical disequilibrium is often cited as a possible means of life detection (Cockell *et al.* 2009; Kasting *et al.* 2009; Léger 2000; Sagan *et al.* 1993; Seager 2014; Seager and Bains 2015; Seager and Deming 2010), and sometimes criticized (Schwartzman and Volk 2004; Seager and Bains 2015). However this idea is not quantified, except in rare and specific instances. For example, Simoncini *et al.* (2013) used kinetic arguments and non-equilibrium thermodynamics to infer the minimum power driving atmospheric disequilibrium for Earth and Mars, and Seager *et al.* (2013) applied kinetic arguments to deduce biomass estimates for biosignature gas detections. Kleidon (2012) reviewed the mechanisms for free energy generation on the Earth and the possible effects of increasing human consumption of free energy, whilst Ulanowicz and Hannon (1987) argued that surfaces dominated by biology such as tropical rainforests are more dissipative than desert surfaces, and that this difference in entropy production might be accessible to remote sensing. Estrada (2012) introduced a novel but perhaps non-intuitive atmospheric disequilibrium metric based on examining the directionality of the network of chemical reactions in an atmosphere. Estrada's method highlights species injected into an atmosphere, but many of them for Earth are anthropogenic, such as halocarbons.

*Thermodynamic* disequilibrium in planetary atmospheres and its quantification for biosignature detection on exoplanets has not been examined for several decades. Lippincott *et al.* (1967) and Lovelock (1975) made early attempts to calculate thermodynamic disequilibrium for the Solar System planets, but knowledge of the actual atmospheric composition of Solar System planets, computational methods, and thermodynamic data for chemical equilibrium calculations have since greatly improved. Additionally, Lovelock (1975), who is the only author to report the magnitude of disequilibrium in Earth's atmosphere-ocean system, did not provide the details of his method. However, we infer that he probably used analytic calculations and assumed that key redox couples reacted to completion (see results section). This method does not give the correct answer for the thermodynamic equilibrium of the Earth atmosphere-ocean system because completion is not necessarily the same as the equilibrium state.

Another important issue is that all atmospheres are in disequilibrium to some extent because they receive a free energy flux from sunlight and, more generally, could obtain additional free energy from release of volcanic gases, tidal energy, or internal heat. Indeed there are already ostensible detections of thermodynamic disequilibrium in the atmospheres of transiting, Jovian-like exoplanets (e.g. Knutson *et al.* (2012); Moses *et al.* (2011); Stevenson *et al.* (2010), although see Line and Yung (2013) for an alternative view). Consequently, inferring life from atmospheric thermodynamic disequilibrium is a question of degree. In order to understand the issue properly, accurate quantification is necessary. Thus, part of the purpose in this work is to examine the abiotic disequilibria in Solar System atmospheres and compare with the Earth.

Here, we present a rigorous methodology and calculation of thermodynamic disequilibrium in the atmospheres of Solar System planets and Saturn's largest moon, Titan, using Gibbs free energy. We quantify chemical disequilibrium in atmospheres as the difference between the Gibbs energy of observed atmospheric constituents and the Gibbs free energy of the same atmosphere if all its constituents were reacted to equilibrium under prevailing conditions of temperature and pressure. For Earth, the purely gas phase calculation does not capture the disequilibrium in the atmosphere-ocean system, and so we present a method for quantifying the atmosphere-ocean disequilibrium using multiphase Gibbs energy minimization. We do not consider kinetic disequilibrium in our analysis, which will be the topic of future work. Finally, we discuss whether using thermodynamic disequilibrium as a biosignature is feasible on both practical and theoretical grounds. To promote cooperation in research, our Matlab source code and all the databases used for these calculations are available as post-publication open source software.

**Methods**

*Gas phase calculations*
Appendix A gives specifics on the gas phase calculations and here we outline the general methodology. To quantify thermodynamic disequilibrium, we model each atmosphere as a well-mixed closed system at a constant pressure (surface pressure or 1 bar for giant planets) and temperature (global mean surface temperature or the mean temperature at the 1 bar level for giant planets). Thermodynamic theory states that for a closed chemical system at constant temperature and pressure, chemical equilibrium is achieved when the Gibbs free energy of the system is minimized.

If there are $N$ chemical species in an atmosphere containing $n_i$ moles of each gas $i$, then the total Gibbs free energy of the system (in Joules) is:

$$G_{(T,P)} = \sum_i^N \left(\frac{\partial G}{\partial n_i}\right)_{T,P} n_i = \sum_i^N \mu_i n_i$$
$$= \sum_i^N \left(\mu_i - G^\circ_{i(T,P_r)} + G^\circ_{i(T,P_r)}\right) n_i \quad (1)$$

Here, $\mu_i$ (J/mol) is the partial molar Gibbs free energy, or equivalently, the chemical potential (see Anderson (2005)), $G^\circ_{i(T,P_r)}$ is the standard partial molar Gibbs free energy of gas $i$ at reference pressure $P_r$ which is typically 1 bar or 1 atm depending on the database used. We have written the expression in the second line of equation (1) because of a basic relationship in thermodynamics (the definition of chemical potential) is:

$$\mu_i - G^{\circ}_{i(T,P_r)} = RT \ln(a_i) = RT \ln\left(\frac{f_i}{f_i^{\circ}}\right) = RT \ln\frac{(P_i \gamma_{fi})}{(P_i \gamma_{fi})^{\circ}} \quad (2)$$

Here, $a_i = f_i/f_i^{\circ}$, is the activity of species $i$, $f_i$ denotes partial fugacity, $f_i^{\circ}$ is a reference partial fugacity, $\gamma_{fi}$ is the activity coefficient of species $i$, with $\gamma_{fi}^{\circ}$ a standard value, and $P_i$ is partial pressure of species $i$. See Anderson (2005, p198-208) for a derivation of equation (2). The activities of reacting species are given by:

$$a_i = \left(\frac{f_i}{f_i^{\circ}}\right) = P_i \gamma_{fi} = \frac{P n_i}{n_T} \gamma_{fi} \quad (3)$$

where we have taken $f_i = 1$ bar as the reference state, $n_i$ is the number of moles of species $i$, $n_T$ is the total number of moles, and $P = \sum_i P_i$ is the total pressure by Dalton's Law.

Substitution of equations (2) and (3) into equation (1) gives the following expression for the Gibbs free energy of an atmosphere, where we drop the '$N$' in the summation to avoid clutter:

$$G_{(T,P)} = \sum_i n_i (G^{\circ}_{i(T,P_r)} + RT \ln(a_i)) = \sum_i n_i (G^{\circ}_{i(T,P_r)} + RT \ln(P n_i \gamma_{fi}/n_T)) \quad (4)$$

This is a form of Gibbs free energy used in previous Gibbs free energy minimization schemes minimum (Eriksson 1971; Eriksson 1975; Venot *et al.* 2013; White *et al.* 1958). Here, $G_{(T,P)}$ is the Gibbs free energy of the system at constant temperature, $T$, and constant pressure, $P$. The number of moles of the $i$-th atmospheric species is given by $n_i$, the standard Gibbs free energy of the $i$-th species at some reference pressure $P_r$ is given by $G^{\circ}_{i(T,P_r)}$, and $R = 8.314$ J/mol is the universal gas constant. Variables $\gamma_{fi}$ and $n_T$ were defined earlier. The summation in equation (4) is over all molecular species in the planet's atmosphere. For purely gas phase calculations we will take $n_T = 1$ so that mixing ratios can be substituted for $n_i$ and all Gibbs energy results will be in units of joules per mole of atmosphere. For example, for Earth, $n_{O_2} \approx 0.21$ and $n_{N_2} \approx 0.78$.

In practice, Gibbs free energy is only defined relative to some reference energy, and so we substitute absolute Gibbs energies for Gibbs energies of formation to obtain:

$$\Delta G_{(T,P)} = \sum_i n_i (\Delta_f G^\circ_{i\,(T,P_r)} + RT \ln(P_i \gamma_{fi})) = \sum_i n_i (\Delta_f G^\circ_{i\,(T,P_r)} + RT \ln(P n_i \gamma_{fi} / n_T))$$

(5)

Here, $\Delta_f G^\circ_{i\,(T,P_r)}$ is the standard free energy of formation for the *i*-th species. This is defined as the free energy change associated with forming the *i*-th species from its constituent elements at temperature $T$ and pressure $P_r$. It can be shown that the minimum of equation (4) is identical to the minimum of equation (5) (see Appendix B for this proof), and so in practice we find the equilibrium state by finding the equilibrium $n_i$ that minimize equation (5), which we represent with an overbar as $\bar{n}_i$.

Temperature dependent standard Gibbs free energies of formation were calculated from enthalpies and entropies of formation retrieved from NASA's thermodynamic database (Burcat and Ruscic 2005). Some atmospheres such as that of Venus have high surface temperature and pressure, and so their constituent gases exhibit non-ideal behavior. We account for this by calculating temperature and pressure dependent fugacity coefficients for each species using the Soave equation as described in Walas (1985, p146). The Soave equation is an empirical equation of state that accounts for the non-zero volume of particles and attractive forces between pairs of particles. To calculate fugacity coefficients for a known mixture of gases at a specified temperature and pressure, the critical temperatures, critical pressures, acentric factors (a measure of non-sphericity of molecules) and binary interaction parameters for all the constituent species are required. We obtained critical temperatures, critical pressures and acentric factors from Perry *et al.* (2008, section 2-136). Tests indicate that binary interaction parameters have a negligible effect on the overall Gibbs energy changes we are interested in, and so all binary interaction parameters were assumed to be zero in our analysis (see Appendix A). Because the fugacity coefficient is a function of species concentration, the fugacity coefficients of all gaseous species were recalculated at every iteration in our optimization routine to ensure convergence to the correct equilibrium. Typically, including fugacity coefficients does not change the results very much for Earth-like temperatures and pressures. However, for high-pressure atmospheres such as Venus, fugacity coefficients are important because the departures from ideal gas behavior are appreciable.

For any observed planetary atmosphere with a composition specified by mole fractions $n_i$, the equilibrium composition can be found by determining the mole fractions $\bar{n}_i$ that minimize $\Delta G_{(T,P)}$ in equation (5) subject to the constraint that atoms are conserved. The atom constraint condition is given by:

*Equilibrium moles of element k = Observed moles of element k*

$$\sum_i v_{ki}\bar{n}_i = \sum_i v_{ki} n_i \tag{6}$$

Here, $v_{ki}$ is the number of atoms of element *k* per molecule of the *i*-th species.

The above framework is a constrained, non-linear optimization problem: the equilibrium state of an atmosphere can be found by minimizing equation (5) subject to equation (6). We used an interior points method (Byrd *et al.* 2000; Byrd *et al.* 1999) implemented using Matlab's *fmincon* function to solve this optimization problem. Interior points is an efficient and reliable optimization technique known to be useful for chemical equilibrium problems (see Karpov *et al.* (1997) and references therein). For gas phase Gibbs energy minimization the equation to be optimized is convex (has non-negative second derivative) and so any local minimum will be the single global minimum (White *et al.* 1958).

To quantify the chemical disequilibrium in a planet's atmosphere, we define the "available Gibbs energy", $\Phi$, as the difference in Gibbs free energy between the observed (initial) state and the equilibrium state:

$$\Phi \equiv G_{(T,P)}(n_i) - G_{(T,P)}(\bar{n}_i) \tag{7}$$

Since Gibbs free energy is only defined relative to some reference energy, in practice we compute available Gibbs energy using this equivalent expression:

$$\Phi = \Delta G_{(T,P)}(n_i) - \Delta G_{(T,P)}(\bar{n}_i) \tag{8}$$

See Appendix B for a proof of the equivalence of equations (7) and (8). The available Gibbs energy, $\Phi$, has units of joules per mole of atmosphere. Thermodynamic theory states that this Gibbs free energy difference is the maximum useful work that can be extracted from the system. In other words, $\Phi$ is the untapped chemical free energy in a planet's atmosphere and so provides our metric of disequilibrium.

*Multiphase calculations*
The numerical approach described above applies to gaseous systems only such as Mars, Venus and Jupiter. To calculate chemical disequilibrium for planets with surface oceans, we reformulate the Gibbs energy expression for multiphase systems. Appendix C gives specifics of the multiphase calculations and here we provide a general overview. We follow Karpov *et al.* (1997), and use the following expression for the Gibbs energy of a multiphase system:

$$\Delta G_{(T,P)} = \sum_i c_i n_i + \sum_\alpha \sum_{i \in \alpha} n_i RT \ln(n_i/n_\alpha) - \sum_{j=aqueous\ species} n_j RT \ln(n_w/n_{aq})$$

$$c_i = \begin{cases} \Delta_f G°_{i\,(T,P)} + RT \ln(\gamma_{fi}) + RT \ln(P), & i \in gas \\ \Delta_f G°_{i\,(T,P)} + RT \ln(\gamma_{aw}), & i \in water \\ \Delta_f G°_{i\,(T,P)} + RT \ln(\gamma_{ai}) + RT \ln(55.5084), & i \in aqueous \end{cases} \quad (9)$$

Here, we have simplified equations in Karpov *et al.* (1997) to exclude solid phases and non-water pure liquids because we don't consider such systems in this study. In addition to the variables already defined above, we have the following:

$\alpha$ = index for the phase (gaseous, water or aqueous).
$n_\alpha$ = total number of moles of species in phase $\alpha$.
$n_w$ = total number of moles of liquid water in the system.
$n_{aq}$ = total number of moles of aqueous species in the system.
$\gamma_{aw}$ = activity coefficient of water.
$\gamma_{ai}$ = activity coefficient of the *i*-th aqueous species.

We see that equation (9) for the gas phase system ($c_i, i \in gas$) is identical to equation (5) if we let $n_\alpha = n_T$.

To calculate the equilibrium state of Earth's atmosphere-ocean system, we minimize equation (9) above subject to the constraint that atoms and charge are conserved, where the latter means that aqueous systems are electroneutral. The atom constraint is identical to that used for the gaseous systems as defined by equation (6). The charge constraint is given by:

*Total charge in equilibrium = Total charge in observed state*

$$\sum_i q_i \bar{n}_i = \sum_i q_i n_i \quad (10)$$

Here, $q_i$, is the charge per molecule of the *i*-th species. Just as for the gaseous calculations, the Gibbs energy difference between the observed and equilibrium states, $\Phi$, can be calculated once the equilibrium state is determined.

Temperature and pressure dependent Gibbs free energies of formation for aqueous species were calculated from the SPRONS96 database in SUPCRT (Johnson *et al.* 1992) and the methodology described in Walther (2009). We assumed that the Born coefficients, which describe species-specific solvation properties, would have a negligible effect on Gibbs energies and so those terms were dropped from the calculations. Activity coefficients for aqueous species were approximated using the Truesdell-Jones equation and thermodynamic coefficients from Langmuir (1997, p133) (see Appendix C). For Earth, the available Gibbs energy is quite sensitive to water activity. Thus rather than use the approximation above, the activity coefficient for water was calculated rigorously using a simplified form of the Pitzer equations (Marion and Kargel 2007) and Pitzer coefficients from Appelo and Postma (2005) and Marion (2002).

Finding the equilibrium state for multiphase systems is more challenging than for single-phase gaseous systems. The Matlab function *fmincon* was once again used to implement the optimization, but this time we provided the analytic first derivative gradient for the Gibbs energy function in equation (9). This ensured more rapid and reliable convergence. For multiphase Gibbs energy minimization problems there is no guarantee that the local minima equal the global minimum (Nichita *et al.* 2002). Consequently, we implemented a simple global minimum search by iterating over a large ensemble of initial conditions and selecting the solution from the ensemble with the minimum of the minima (see Appendix C for details).

*Semi-analytic validation*
Validation of gas phase calculations was initially done using a classic textbook case from Balzhiser *et al.* (1972, p. 513-527) which was found to match our numerical calculations. This textbook case was a gas phase reaction of ethane and steam at 1000K to form $H_2$, CO and various hydrocarbons. We also correctly solved the equilibrium using the method of Lagrange multipliers as a check.

Furthermore, to corroborate the numerical Gibbs free energy calculations for planetary atmospheres, we also approximated the available Gibbs energy in each atmosphere using a simple analytic expression. For a single reaction between arbitrary reactants and products,

$$AR_1 + BR_2 \rightleftharpoons CP_1 + DP_2 \tag{11}$$

The Gibbs energy of this reaction is given by:

$$\Delta_r G = \Delta_r G^\circ + RT \ln(Q) = \Delta_r G^\circ + RT \ln\left(\frac{a_{P1}^C a_{P2}^D}{a_{R1}^A a_{R2}^B}\right) \tag{12}$$

Here, A, B, C and D are the stoichiometric coefficients representing reactants $R_1$, $R_2$, products $P_1$ and $P_2$, respectively. The activity of each species is $a_X$, $R$ is the universal gas constant and $T$ is the temperature of the system. The left hand side of equation (12), $\Delta_r G$, is the change in Gibbs energy of the system per *A* moles of reactant $R_1$ and B moles of reactant $R_2$ that are converted to products. The standard free energy of the reaction, $\Delta_r G^\circ$, represents the Gibbs energy of the reaction when the activities of all species equal unity. At equilibrium, the left hand side of equation (12) equals zero. This equilibrium can be found by appropriate substitution for each of the activities in terms of initial abundances and the total moles reacted to reach equilibrium (the only unknown variable), and by solving the resultant polynomial (see Appendix D). This equilibrium condition is equivalent to minimizing the Gibbs energy of the same system using equation (5).

The available Gibbs energy of the system, $\Phi$, can be obtained by integrating $\Delta_r G$ from the initial state to the equilibrium state. Strictly speaking, this semi-analytic approach can only be applied to systems of gases where there is only one possible reaction, and not to complex mixtures of gases such as planetary atmospheres. However, this calculation can be repeated for all the key reactions in a planet's atmosphere, and the summed available Gibbs energies can be compared to the numerical result from Gibbs energy minimization. The key reactions for these semi-analytic calculations were chosen using the important redox couples identified by chemical intuition for each atmosphere. The two approaches are not exactly equivalent because treating each reaction independently does not account for interactions between multiple reactions. To simplify the semi-analytic calculations we also make the assumption that the total moles in the atmosphere remained unchanged as the reaction proceeds. Consequently we expect small differences between the semi-analytic and numerical approaches. Appendix D gives step-by-step detail on semi-analytic procedures. The semi-analytic approximations described above for gas phase systems can also be applied to aqueous reactions in a multiphase system such as the Earth.

*Validation using Aspen Plus*
Both gaseous and multiphase calculations were validated using the commercial software package Aspen Plus (Version 8.6), which is commonly used in chemical engineering. Aspen Plus provided a completely independent check of our calculations because it uses different thermodynamic databases and property models to both our Matlab calculations and semi-analytic approximations. We used an equilibrium reactor called "RGIBBS" in Aspen Plus to implement gas phase and multiphase equilibrium calculations by Gibbs free energy minimization. We also used the Peng-Robinson equation of state (Prausnitz *et al.* 1999) model for gas phase calculations, which is appropriate for the temperatures and pressures we are interested in. For multiphase calculations, we used a "Flash2" phase separator in the Aspen Plus model in addition to an RGIBBS reactor, which ensured that the phases of aqueous species were correctly assigned. Without the phase separator, the equilibrium results were unphysical, and the resultant Gibbs energy change was inaccurate. We report results from the Electrolyte Non-Random Two Liquid (ELECNRTL) model in the main text (e.g. see Prausnitz *et al.* (1999) chapter 6). ELECNRTL is the recommended activity coefficient model for calculations involving electrolytes (Aspen Technology Inc.

2000). Appendix E explains the Aspen Plus multiphase calculation in more detail, and reports results for a different electrolyte model.

*Planetary data*

The observed atmospheric compositions used in this analysis were obtained from a variety of up-to-date sources. The atmospheric composition of Venus at the surface was taken from Fegley (2014, p. 131) and Krasnopolsky and Lefèvre (2013, p. 64). The atmospheric composition of Mars at the surface was taken from Lodders and Fegley (1998) but updated with *Curiosity Rover* observations (Baines *et al.* 2014; Mahaffy *et al.* 2013). The atmospheric composition of Jupiter at 1 bar was taken from Lodders and Fegley (1998) but updated using the compilation in Irwin (2009, p. 100-3). The atmospheric composition of Titan at the surface was taken from the review by Catling (2015). Uranus' atmospheric composition at 1 bar was inferred from the review by Irwin (2009, p. 124), Lodders and Fegley (1998) and Catling (2015). Finally, Earth's atmospheric composition was assumed to be that of the US Standard Atmosphere and the abundance of dissolved ions in average seawater was obtained from Pilson (2012, p. 59). Nitrate abundance was obtained from Gruber (2008, p. 13).

**Results**

Tables 1-7 show equilibrium calculations for the Solar System atmospheres. The format of each table is the same: the first column lists the species present in each body's atmosphere, the second column gives the observed mixing ratios of these species, and the third column is the species abundances at equilibrium, as determined by our Gibbs free energy minimization code. The fourth column is an independent validation of our calculations where the equilibrium abundances are determined using the commercial software package, Aspen Plus. The equilibrium abundances from our Gibbs energy minimization and from Aspen Plus match very closely in every case. Bolded rows highlight the species where abundances change during the reaction to equilibrium. Figures 1-7 are the graphical representation of tables 1-7, respectively. Observed (black bars) and equilibrium (grey bars) abundances of all species for each atmosphere are plotted on a log scale. We only plot the equilibrium abundances from our Gibbs energy minimization calculations and not from Aspen Plus since the differences are barely visible. All equilibrium calculations are performed at observed mean surface temperature and pressure conditions (for terrestrial planets), or at 1 bar and the mean temperature at 1 bar (for giant planets with no surface), unless stated otherwise.

Note that although the observed abundances in the tables and figures are mixing ratios, the equilibrium abundances do not sum to exactly unity. The equilibrium molar abundances are instead the moles of each species that remain when 1 mole of the observed atmosphere reacts to equilibrium (reaction to equilibrium conserves atoms but does not conserve the number of moles in an atmosphere). We chose not to renormalize the equilibrium abundances to obtain mixing ratios because it was easier to identify which species are involved in reactions from the tables without normalization.

Table 8 shows the available Gibbs energy, $\Phi$ (defined in equation (7)) in each planet's atmosphere, and figure 8a is a graphical representation of these results. The second column in table 8 gives the available Gibbs energy as determined by our own numerical code for Gibbs energy minimization. Column 3 shows the semi-analytic approximation of available Gibbs energy that were calculated by choosing key reactions, finding their equilibria independently, and summing the Gibbs energy changes associated with each reaction (see methods section). Column 4 is the available Gibbs energy in each atmosphere as determined by the commercial software package, Aspen Plus. In almost every case, the available Gibbs energies from these three methods are consistent to within a few percent or better. The excellent agreement is encouraging because Aspen Plus uses different thermodynamic databases and models to our Gibbs energy minimization calculations. Column 5 gives the Gibbs free energy of each planet as reported by Lovelock (1975). The discrepancies between their results and our results are attributable to much improved knowledge of atmospheric compositions and our more accurate computational techniques (see below).

We now discuss what accounts for the disequilibrium in each atmosphere. We do this by identifying the key gases that are in disequilibrium and describing how the chemical conditions on each body give rise to various disequilibria.

*Venus*
The disequilibrium in Venus' lower atmosphere is comparatively small, which is expected because the high pressure and temperature favors chemical reactions that push the atmosphere close to equilibrium (Yung and DeMore 1999, p292). There is little difference between the observed abundances and equilibrium abundances (figure 1, table 1) except for very minor species. Consequently, the available Gibbs energy in Venus' atmosphere is only ≈0.06 J/mol (table 8).

The largest contributor to the disequilibrium in Venus' atmosphere (in terms of available energy) is the coexistence of elemental sulfur (S) and carbon dioxide ($CO_2$). Semi-analytic calculations predict that the following reaction should deplete all the elemental sulfur in Venus' atmosphere:

$$2CO_2 + S \rightleftharpoons SO_2 + 2CO \tag{13}$$

Gibbs energy minimization calculations confirm that elemental sulfur is absent in equilibrium. The disequilibrium in Venus' atmosphere is maintained by photochemistry; photochemical dissociation of $SO_2$ and OCS in the upper atmosphere maintains out of equilibrium sulfur chemistry (Yung and DeMore 1999, p. 292).

*Mars*

The disequilibrium in Mars' atmosphere is large compared to other Solar System atmospheres. The available Gibbs energy in Mars' atmosphere, 136 J/mol, is 1-2 orders of magnitude greater than every other atmosphere we consider except for Earth's atmosphere-ocean system. Figure 2 and table 2 show several abundant constituents in Mars' atmosphere with observed mixing ratios substantially different from equilibrium abundances.

The largest contributor to disequilibrium in Mars' atmosphere (in terms of available energy) is the coexistence of CO and $O_2$. Both semi-analytic and numerical calculations predict that, in equilibrium, all the CO should be oxidized by $O_2$ to form $CO_2$ by the following reaction:

$$2CO + O_2 \rightleftharpoons 2CO_2 \tag{14}$$

This is confirmed by the stoichiometry of the change in abundances from the numerical calculation (column 5, table 2). Reaction to equilibrium decreases the abundance of $O_2$ by $2.8538 \times 10^{-4}$ moles and decreases the abundance of CO by $5.57 \times 10^{-4}$ moles, i.e. almost a 1:2 ratio. The abundance of $CO_2$ increases by $5.57 \times 10^{-4}$ moles. The stoichiometry is not exactly the same as equation (14) because oxygen is also depleted by reaction with hydrogen gas by the following reaction:

$$2H_2 + O_2 \rightleftharpoons 2H_2O \tag{15}$$

Since numerical calculations indicate that hydrogen decreases by $0.15 \times 10^{-4}$ moles (column 5, table 2), this implies molecular oxygen must decrease by $0.15 \times 10^{-4}/2 = 0.075 \times 10^{-4}$ moles. Subtracting this decrease from the overall change in oxygen yields $(2.8538-0.075) \times 10^{-4} = 2.7788 \times 10^{-4}$ moles, which is a closer match to the stoichiometry in equation (14). The remaining discrepancy is similarly explained by the reaction of ozone to form molecular oxygen:

$$2O_3 \rightleftharpoons 3O_2 \tag{16}$$

Disequilibrium in Mars' atmosphere is maintained by photochemistry. The photodissociation of $CO_2$ continuously replenishes CO in the Martian atmosphere (Nair *et al.* 1994; Zahnle *et al.* 2008). The Martian atmosphere also has an overabundance of $H_2$ and $O_3$, both of which are maintained by photodissociation of water.

The difference between the available Gibbs energy in Mars' atmosphere and the available energy in other photochemically driven atmospheric disequilibria can be partly explained by differences in atmospheric column mass and chemical complexity. Since Mars' atmosphere is more tenuous than other atmospheres, and lacks species (e.g., chlorine-bearing gases) that enable more pathways of catalytic recombination of $CO_2$, photochemical reactions have a greater effect on overall composition. In contrast, photochemistry on Venus does not result in large available Gibbs energy per mole of atmosphere because the thick atmosphere along with efficient catalysts buffers its effect on composition. The $CO_2$ column photodissociation rates on Mars and Venus are comparable, $2\times10^{12}$ molecules/cm$^2$/s (Huguenin *et al.* 1977) and $7.6\times10^{12}$ molecules/cm$^2$/s (Bougher *et al.* 1997, p. 448), respectively, whereas the $SO_2$ column photodissociation rate, or equivalently the $H_2SO_4$ production rate, on Venus is ~$5.6\times10^{11}$ molecules/cm$^2$/s (Krasnopolsky 2015; Zhang *et al.* 2012) . However, the column mass is larger on Venus. The column mass is $P/g$, where $P$ is surface pressure and $g$ is gravitational acceleration. On Venus, the column mass is $93.3\times10^5$ Pa/ 8.87 m/s$^2$ = 1,051,680 kg/m$^2$ (taking the pressure at the mean elevation), whereas on Mars the column mass is 600 Pa/ 3.711 m/s$^2$ = 159.1 kg/m$^2$, so the Venus:Mars ratio is 1,051,680/159.1 ~ 6,600. Whereas Mars has catalytic recombination of $CO_2$ from only odd hydrogen species, Venus has more efficient catalytic cycles involving Cl, N and H species for $CO_2$ recombination (Yung and DeMore 1999, p. 249,288), such that $O_2$ on Venus has an upper limit concentration <0.3 ppmv. Consequently, the products of $CO_2$ dissociation do not significantly influence the disequilibrium on Venus; instead, sulfur chemistry makes the dominant contribution, as discussed earlier. The net result is that the available free energy in Venus' atmosphere is ~2000 times smaller than that of Mars (table 8).

Note that Mars' atmospheric composition varies seasonally via $CO_2$ exchange with polar caps, and on longer timescales obliquity cycles will modulate atmospheric $CO_2$ due to regolith adsorption. However, these changes in atmospheric composition are unlikely to have a large effect on the available energy in Mars' atmosphere. The total $CO_2$ reservoir in the regolith and the polar ice is equivalent to 5-30 mbar $CO_2$ (Covey *et al.* 2013, p. 171). Zahnle *et al.* (2008) used a 1D photochemical model to compute self-consistent Mars atmospheres with $pCO_2$ varying from 1-100 mbar, thereby encompassing the range of atmospheric variability from seasonal and obliquity variations. We computed the available energy for this range of photochemical outputs and found that it was less than 200 J/mol regardless of pCO$_2$. Although the photochemical model calculates the water volume mixing ratio from a specified relative humidity, H$_2$O is redox neutral and so changing its abundance will not have a large effect on available energy.

*Jupiter*

The disequilibrium in Jupiter's atmosphere at the 1 bar level is very small compared to other atmospheres in the Solar System (≈0.001 J/mol). The observed mixing ratios and equilibrium abundances (table 3, figure 3) are virtually identical; the largest changes are at the parts per billion level. The small disequilibrium in Jupiter's atmosphere is attributable to the coexistence of HCN with H$_2$ and the coexistence of CO with H$_2$. Both numerical and semi-analytic calculations predict that HCN and CO should be completely depleted in equilibrium by the following reactions

$$3H_2(g) + HCN(g) \rightleftharpoons CH_4(g) + NH_3(g) \qquad (17)$$

$$3H_2 + CO \rightleftharpoons CH_4 + H_2O \qquad (18)$$

This is confirmed by the stoichiometry of the change in abundances from the numerical calculation (column 5, table 3): HCN and CO abundances decrease by 3.6×10$^{-9}$ and 1.6×10$^{-9}$ moles, respectively, whereas NH$_3$ and H$_2$O abundances increase by 3.6×10$^{-9}$ and 1.6×10$^{-9}$ moles, respectively. Based on these numbers and equations (17) and (18), we would predict that CH$_4$ abundance should increase by (3.6+1.4)×10$^{-9}$ = 5.2×10$^{-9}$ moles, and that H$_2$ abundance should decrease by 3×(3.6×10$^{-9}$ + 1.6×10$^{-9}$) = 1.56×10$^{-9}$ moles. These predictions exactly match the observed changes in these species in table 3.

It is not surprising that Jupiter's atmosphere is very close to equilibrium. Photochemically produced disequilibrium species are vigorously mixed into the high temperature interior (1000 K) where they are hydrogenated to reform equilibrium species (Lewis 2012, p. 209-212). The small disequilibrium that remains is attributable to a combination of deeper vertical mixing, material delivery, and photochemistry. CO is thermodynamically favored in the very high-temperature interior and deep vertical mixing delivers it to the upper atmosphere (Prinn and Barshay 1977), although some infall of material from space is required to explain observed CO abundances (Bézard *et al.* 2002). HCN is also thermodynamically favored in the interior, but observed abundances are best explained by photochemical sources (Kaye and Strobel 1983).

We repeated the equilibrium calculation for Jupiter at the 1 millibar level. This is of interest for exoplanet characterization since infrared spectroscopy may be limited to probing the millibar level of Jovian-like atmospheres due to thick clouds or hazes. The mean temperature at 1 millibar is approximately equal to the temperature at 1 bar due to the temperature inversion in Jupiter's stratosphere, and consequently any difference in available Gibbs energy can be ascribed to changing mixing ratios. Using stratospheric species abundances from Irwin (2009, p. 101), we found the available Gibbs energy in Jupiter's atmosphere at 1 millibar to be 0.35 J/mol. This disequilibrium can be ascribed to photochemically replenished organics such as $C_2H_6$, and to a lesser extent $C_2H_2$ and $C_2H_4$.

*Titan*
The moderate disequilibrium in Titan's atmosphere (≈1.2 J/mol) is also driven by photochemistry. Both ethane ($C_2H_6$) and acetylene ($C_2H_2$) exist in Titan's atmosphere, whereas semi-analytic and numerical calculations predict that these species should be depleted in equilibrium by reactions with $H_2$ to form $CH_4$:

$$C_2H_6 + H_2 \rightleftharpoons 2CH_4 \quad (19)$$
$$C_2H_2 + 3H_2 \rightleftharpoons 2CH_4 \quad (20)$$

Table 4 confirms this stoichiometrically: ethane and acetylene abundances decrease by $1\times10^{-5}$ and $2\times10^{-6}$, respectively, by reaction to equilibrium (table 4). From the equations above, this would imply that $CH_4$ abundance should increase by $2\times (1\times10^{-5}+2\times10^{-6}) = 2.4\times10^{-5}$ whereas hydrogen abundance should decrease by $3\times2\times10^{-6}+1\times10^{-5}=1.6\times10^{-5}$. These predictions exactly match the observed changes in these species in table 4.

We have not included Titan's hydrocarbon lakes in this calculation for several reasons. Firstly, the thermodynamics of cold hydrocarbon solutions is beyond the scope of this study and poorly known. Second, the composition of lakes on Titan and the degree to which they are variable are unknown. Third, current hypothetical estimates of lake composition (Cordier *et al.* 2009; Glein and Shock 2013) are based on purely thermodynamic equilibrium models and so are inappropriate for revealing disequilibrium. The total volume of Titan's lakes is estimated to be 32,000 km$^3$ (Lorenz et al. 2014). If we assume the lake density is 654 kg/m$^3$, which is the density of liquid ethane at 92.5 K (Younglove and Ely 1987), then the total mass of the lakes is $2.1\times10^{16}$ kg. The surface pressure on Titan is 1.5 bar, the surface area is $8.3\times10^7$ km$^2$, and the surface gravity is 1.35 ms$^{-2}$. This implies the mass of the atmosphere is $(8.3\times10^7\times1000^2\ m^2)\times(10^5\times1.5\ Pa)/(1.35\ ms^{-2}) = 9.2\times10^{18}$ kg. Thus, the mass of the lakes is 0.2% the mass of Titan's atmosphere. Disequilibrium species at the parts per thousand level can impact the available energy, as evidence by the CO-$O_2$ pairing in Mars' atmosphere. Consequently, if Titan's lakes are in chemical disequilibrium with the atmosphere, then the available Gibbs energy of the total fluid reservoir may be larger than the atmosphere-only result we report here.

*Uranus*

Observational knowledge of Uranus' atmosphere is limited and so it is difficult to calculate disequilibrium at the 1 bar level. Table 5a shows the observed abundances at 1 bar; there is insufficient diversity of molecular species for any reactions to occur. The observed composition is the same as the equilibrium composition, and the available Gibbs energy in Uranus' troposphere is nominally 0 J/mol. In reality, there are probably trace species at 1 bar that contribute to a small disequilibrium. To place an upper bound on the disequilibrium in Uranus' atmosphere, we included trace species from the stratosphere in our calculations (Table 5b). Of course, the stratosphere for most planets with thick atmospheres and shortwave stratospheric absorbers is located vertically above the ~0.1 bar level (Robinson and Catling 2014) and so is not at the 1 bar level that we use for Gibbs energy calculations. Even so, when semi-analytic and numerical calculations are repeated for this case, we find the maximum disequilibrium in Uranus' atmosphere is still comparatively small, 0.097 J/mol.

*Saturn and Neptune*
Saturn and Neptune were excluded from our analysis because of their close similarity to Jupiter and Uranus, respectively. Essentially, in Jupiter and Uranus, we chose a representative of the gas and ice giants, respectively.

*Earth*
We calculated the disequilibrium in Earth's atmosphere for two different cases. Firstly, we considered only the Earth's atmosphere (figure 6, table 6). There are minor differences between the observed atmospheric composition and the equilibrium composition. The largest contributor to gas phase disequilibrium in Earth's atmosphere is the coexistence of $O_2$ and $CH_4$, and the available Gibbs energy in the atmosphere is only 1.5 J/mol, which is not unusual compared to other Solar System atmospheres. The $O_2$ and $CH_4$ couple contributes ~90% of this gas phase disequilibrium (1.3 J/mol).

Next, we consider the multiphase equilibrium calculation that includes the Earth's atmosphere and oceans with dissolved ion species (figure 7, table 7). In this case, the disequilibrium in Earth's atmosphere-ocean system is very large; the available Gibbs energy is 2326 J/mol of atmosphere.

The large disequilibrium is attributable to the coexistence of $N_2$, $O_2$ and liquid water. Both numerical and semi-analytic calculations predict that these three species should react to form nitrate and hydrogen ions according to the following reaction:

$$2N_2(g) + 5O_2(g) + 2H_2O(l) \rightleftharpoons 4H^+(aq) + 4NO_3^-(aq) \qquad (21)$$

In equilibrium, most of the oxygen in Earth's atmosphere reacts to form hydrogen ions and nitrate (table 7). It has been known for many decades that the coexistence of $N_2$, $O_2$, and $H_2O$ is the largest contributor to disequilibrium in Earth's atmosphere ocean system (Hutchinson 1954, p399; Lewis and Randall 1923; Lovelock 1975; Sillén 1966). However, this is the first time the free energy associated with that disequilibrium has been accurately calculated. Lovelock (1975) reported that the free energy in Earth's atmosphere-ocean system was $5.5 \times 10^4$ J/mol (table 8), which is over an order of magnitude larger than our result. He did not describe their methodology, but we suspect that he assumed the Gibbs energy of the $N_2$-$O_2$-$H_2O$ reaction does not change as the reaction proceeds, and simply multiplied the Gibbs energy of the reaction (at observed abundances) by the number of moles of oxygen in Earth's atmosphere. This approach also assumes the reaction goes to completion with total $O_2$ removal rather than equilibrium. Preliminary analyses by the authors of this study (Catling and Bergsman 2009; Catling and Bergsman 2010) reached a similar result using this methodology. Both the semi-analytic and numerical calculations in this study account for the fact that the Gibbs energy of the reaction diminishes rapidly as oxygen in the atmosphere is depleted, and so the available Gibbs energy in Earth's atmosphere-ocean system, 2326 J/mol, is smaller than previously reported.

Of course, the equilibrium metric is only a hypothetical way of assessing untapped free energy. In reality, $O_2$ also reacts with surface minerals (oxidative weathering) and would be even more depleted with additional free energy if solids were included in the equilibrium model. But we restrict ourselves to gas and gas-liquid equilibrium because those are tractable ways of comparing planets that are tied to quantities that can be observed remotely (see more in discussion section).

We confirmed that the available Gibbs energy in Earth's atmosphere-ocean system is attributable to reaction (21) by repeating the multiphase calculation but excluding $H^+$ and $NO_3^-$. In this case the available Gibbs energy is only 6 J/mol. If we only include the five most important species ($N_2$, $O_2$, $H_2O$, $H^+$ and $NO_3^-$) in multiphase equilibrium calculations, then the available Gibbs energy change is 1812 J/mol (note this includes the effects of changing water activity – see below for details). The difference between this and the total available energy for the Earth system is attributable to carbon-bearing species.

The dissolution of hydrogen ions and nitrate in the ocean by equation (21) acidifies the ocean, which affects the carbonate-$CO_2$ system. By Le Châtelier's principle, as the ocean is acidified, carbon in the form of carbonate and bicarbonate ions converts to atmospheric $CO_2$ and dissolved $CO_2$:

$$CO_2 + H_2O \rightleftharpoons H_2CO_3 \rightleftharpoons H^+ + HCO_3^- \rightleftharpoons 2H^+ + CO_3^{-2} \qquad (22)$$

This reaction shifts to the left as the concentration of hydrogen ions is increased. The Gibbs energy change associated with this shift adds to the overall Gibbs energy change in the Earth's atmosphere-ocean system. Additionally, the dissolution of nitrate and hydrogen ions in water decreases the water activity, which further contributes to the overall Gibbs energy change. If water activity is held fixed, then the available Gibbs energy for Earth is around 600 J/mol less than if water activity is included.

Validating our results for the Earth atmosphere-ocean system is more complex than for gas phase systems. This is because the semi-analytic method we have adopted does not account for the decrease in water activity due to increased nitrate and hydrogen ion abundances. Rather than attempt to compute water activities analytically, we calculated the Gibbs energy change associated with equation (21) in isolation, and compared this to the numerical Gibbs energy minimization calculation using only the five species in this reaction and with water activity fixed to equal 1. The available Gibbs energies for these two cases are shown in table 9; the two values agree to within 1%. Next, we computed the available energy semi-analytically using both reaction (21) and two key reactions that involve carbon-bearing species (see appendix D). This was compared to the numerical Gibbs energy minimization calculation for the same set of species with the water activity set equal to 1. In this case, the available Gibbs energy values also agreed to within 1% (table 9). The difference between this result and the complete numerical calculation can be explained by the effect of water activity. In the numerical calculations the water activity decreases from 0.981877 to 0.981284 from observed to equilibrium state. Following equation (9), this corresponds to a change in Gibbs free energy of:

$$\begin{aligned}\Delta G &\approx n_w RT \left( \ln \gamma_1 - \ln \gamma_2 \right) \\ &= 436 \times 8.314 \times 288.15 \left( -0.01828897 + 0.01889366 \right) \\ &= 631 \text{ J/mol} \end{aligned} \qquad (23)$$

Here, $n_w$ = 436 moles of $H_2O(l)$ per mole of atmosphere, which is derived from the moles of $H_2O$ in the ocean ($7.67 \times 10^{22}$ = $1.38 \times 10^{21}$ kg/(0.018 mol $H_2O$/kg)) and moles of air ($1.76 \times 10^{20}$) as their ratio, 436 = $7.67 \times 10^{22}/1.76 \times 10^{20}$. The value of 631 J/mol is approximately equal to the difference between the numerical calculation including carbon species (water activity=1) and the full numerical calculation (2326 − 1716 = 610 J/mol).

We conclude that the total available Gibbs energy of the Earth atmosphere-ocean, 2326 J/mol, can be explained almost completely by the nitrate reaction (1059 J/mol), the change in carbon-bearing species due to ocean acidification (657

J/mol), and the associated change in water activity (610 J/mol). This conclusion is supported by both numerical and semi-analytic calculations. We do not account for the pressure decrease in our Gibbs energy calculations from depleting the atmosphere of oxygen given that Gibbs energy is defined for a system at constant pressure and temperature. Our multiphase calculations for the Earth should be treated as a constant-pressure approximation.

Although the coexistence of $O_2$ and $CH_4$ is the largest contributor to disequilibrium for a calculation of Earth's atmosphere excluding the oceans, this pair provides a small contribution to the disequilibrium in the total atmosphere-ocean system. If methane is excluded from the Earth atmosphere-ocean equilibrium calculation, then the available Gibbs energy changes from 2325.76 J/mol to 2324.46 J/mol. Similarly, semi-analytic calculations for the reaction, $2O_2 + CH_4 \rightleftharpoons CO_2 + 2H_2O$ yield a Gibbs energy change of only 1.3 J/mol. Methane doesn't contribute much to thermodynamic disequilibrium because of its low abundance of 1.7 ppmv (in the US Standard Atmosphere, noting that anthropogenic emissions mean that the current mean global abundance is slightly higher at ~1.8 ppmv). This doesn't imply that the $O_2$-$CH_4$ disequilibrium is unimportant for life detection purposes. A compelling argument for biogenic fluxes can be made from the coexistence of $O_2$ and $CH_4$ in Earth's atmosphere based on kinetic lifetimes. However, the $O_2$-$CH_4$ pairing is not an important contributor to the available Gibbs energy of thermodynamic disequilibrium in Earth's atmosphere-ocean system.

To express available Gibbs energy as a dimensionless metric, figure 8b plots the available Gibbs free energy in each planet's atmosphere normalized by *RT*, where *T* is the mean temperature for each planet. The value *RT* is the molar thermal energy and depends on solar flux along with Bond albedo and greenhouse effect. Thus, the normalization is a first order and rough correction for the fact that the inner planets receive more free energy input from the Sun that can drive disequilibrium. Figure 8b is similar to figure 8a because surface or 1 bar temperatures vary by an order of magnitude at most, whereas the available Gibbs energies vary by many orders of magnitude. In figure 8b the Earth stands out as only planet in Solar System with chemical disequilibrium comparable in magnitude to thermal energy.

**Discussion**
*Interpretation of thermodynamic disequilibrium*
Earth is unique in the Solar System as the only planet with both a large disequilibrium in its atmosphere-ocean system and a productive surface biosphere (with the caveat that we have not included Titan's lakes in our calculations). This disequilibrium is maintained by life. Atmospheric $O_2$ is produced almost exclusively by oxygenic photosynthesis, and atmospheric $N_2$ is regulated by bacterial nitrification and denitrification. Were denitrification to shut off and biologic N fixation left to operate, the $N_2$ lifetime would be ~10 myr (Jacob 1999, chapter 6).

In the absence of any biogenic fluxes or geological oxygen sinks such as oxidative weathering, reaction (21) would proceed slowly due to lightning, eventually depleting the atmosphere of oxygen (there is some abiotic denitrification but the flux is very small compared to biological denitrification (Devol 2008)). The modern rate of production of nitrogen oxide radicals NOx (=NO + $NO_2$) from lightning is 2-20 Tg(N)/year (Rakov and Uman 2007). OH radicals or ozone further oxidize NOx species intro nitrate that ends up on Earth's surface. Given that the mixing ratio of $N_2$ decreases by approximately 0.08 in our equilibrium calculations, and that the total number of moles of air in the atmosphere is $1.76 \times 10^{20}$, this implies that $1.408 \times 10^{19}$ moles of $N_2$ ($2.816 \times 10^{19}$ moles of N) are converted to nitrate by reaction to equilibrium. Therefore, it would take approximately $2.816 \times 10^{19} / (2\text{-}20 \times 10^{20}/14) = $ 20-200 million years for atmospheric oxygen to be depleted by lightning and converted to nitrate. The coexistence of oxygen, nitrogen and liquid water in the Earth's atmosphere-ocean system is thus evidence a biosphere acting over geologic timescales. Capone *et al.* (2006) also noted that denitrification sustains atmospheric nitrogen on Earth, and that although there are abiotic pathways that deplete $N_2$ (namely lightning), nitrates are not easily converted back to $N_2$ abiotically. In contrast, Kasting *et al.* (1993) argued than on the prebiotic Earth, most of the Earth's nitrogen would reside in the atmosphere in steady state. This is because nitrate is reduced to ammonia in mid-ocean ridge hydrothermal systems, which may then return to the atmosphere and be photochemically converted back to $N_2$. In practice, however, the reduction of nitrate will also yield ammonium (Bada and Miller 1968; Smirnov *et al.* 2008), which will be subsequently sequestered into clay minerals and thereby removed from the atmosphere-ocean reservoir (Summers *et al.* 2012).

It is worth considering why this large disequilibrium exists in Earth's atmosphere-ocean system, and whether we would expect other biospheres to generate large disequilibria. In some respects Earth's large disequilibrium is surprising since life typically exploits environmental free energy gradients rather than generate them. In fact, the $O_2$-$N_2$-water disequilibrium is an incidental byproduct of oxygenic photosynthesis. In addition to producing molecular oxygen, oxygenic photosynthesis also produces large quantities of organic carbon that is buried in sediments. Despite ongoing nitrification and the thermodynamic favorability of reaction (21), nitrate does not accumulate and deplete the atmosphere of oxygen.

This is because denitrifying microbes in anoxic sediments exploit the redox gradient that exists between reduced organic carbon and nitrate (Devol 2008). Without oxygenic photosynthesis producing both $O_2$ and reduced carbon, Earth's atmosphere-ocean disequilibrium would not persist.

In our calculated equilibrium for the Earth atmosphere-ocean system, the molar abundance of $H^+$ ions is 0.14 moles per mole of atmosphere, which corresponds to an ocean pH of 1.7. Lewis and Randall (1923, p567-568) recognized that the equilibrium state of the Earth's atmosphere-ocean system would be highly acidic: "Even starting with water and air, we see... that nitric acid should form... until it reaches a concentration... where the calculated equilibrium exists. It is to be hoped that nature will not discover a catalyst for this reaction, which would permit all of the oxygen and part of the nitrogen of the air to turn the oceans into dilute nitric acid". However, the low pH equilibrium state that we obtain is unlikely to be the state actually realized if life disappeared from the Earth, volcanic fluxes ceased, and the system relaxed to equilibrium. In practice, acidic ocean pH from nitrate dissolution would be buffered by reaction with the crust, for instance by delivery of cations from continental weathering or weathering of seafloor basalt. Nevertheless, we have done calculations where the ocean pH is buffered to pH 8.2, and the Gibbs free energy from reaction to equilibrium is several times larger than our original result.

This discussion highlights the point that we have not included any interactions with solid states of matter in our equilibrium calculations. If the Earth's atmosphere-ocean system were allowed to relax to equilibrium then almost all of the atmospheric $O_2$ would react with the crust via oxidative weathering. There would be a large Gibbs energy change associated with this crustal oxidation. Additionally, there is a large disequilibrium between organic carbon and ferric iron in the crust, both of which have accumulated over time because of photosynthesis and the escape of hydrogen to space (Catling *et al.* 2001). Although there are $3.7 \times 10^{19}$ moles of $O_2$ in the atmosphere and oceans, there are $5.1 \times 10^{20}$ moles $O_2$ equivalent $Fe^{3+}$ and sulfate in sedimentary rocks, and $1.6$-$2.5 \times 10^{21}$ moles $O_2$ equivalent excess $Fe^{3+}$ in igneous and high grade metamorphic rocks (Catling *et al.* 2001; Hayes and Waldbauer 2006; Sleep 2005). These crustal oxidants are in disequilibrium with the $\leq 1.3 \times 10^{21}$ moles $O_2$ equivalent reduced organic carbon in the crust (Catling *et al.* 2001; Wedepohl 1995). Thus, we expect the disequilibrium in Earth's entire crustal reservoir to be several orders of magnitude greater than the disequilibrium in the atmosphere-ocean system in isolation.

However, we chose to exclude interactions with solid phases since we are interested in disequilibrium as a biosignature for exoplanets; the composition of exoplanet crustal material will not be accessible to remote sensing. Instead, our available Gibbs energy metric captures disequilibrium in the gaseous and aqueous phases. In principle, atmospheric composition and the presence of an ocean can be inferred from future telescope observations.

The thermodynamic biosignature metric described in this paper is complementary to kinetic biosignature metrics concerning the fluxes and timescales of gases that should quickly react, such as coexisting oxygen and methane. For example, if the atmospheric abundances of oxygen and methane can be observed, and the abiotic sinks for oxygen and methane can be estimated, then the magnitude of source fluxes required to maintain steady state can be estimated. Biogenic processes may be invoked if these source fluxes are implausibly large compared to all known abiotic sources of oxygen and methane. We calculate that in thermodynamic chemical equilibrium, all $CH_4$ would be absent from the Earth's atmosphere (see results). The average lifetime of a $CH_4$ molecule from photochemical models is ~10 years, and so we can deduce an estimate of the required $CH_4$ flux. For consumption of 1.7 ppmv $CH_4$ in 10 years: $(1.7 \times 10^{-6}) \times (1.8 \times 10^{20}$ moles air) / (10 years) = $3.1 \times 10^{13}$ moles $CH_4$/year flux. The magnitude of this flux is large and on the Earth is dominantly (~90%) biogenic (Kirschke *et al.* 2013).

These flux arguments can be extended to estimate the surface biomass (Seager *et al.* 2013), or the minimal driving power (Simoncini *et al.* 2013) required to maintain steady-state atmospheric abundances. These estimates could provide additional insight into whether the observed disequilibrium is plausibly biogenic in origin: for instance, if the biomass estimate is unreasonably large, or if the driving power is comparable to abiotic processes, the inference to biology is weakened.

It should be noted that our gas phase calculations for the Earth are entirely consistent with the minimal driving power calculations in Simoncini *et al.* (2013). We determined that the available energy from the $CH_4$ and $O_2$ reaction in Earth's atmosphere is 1.3 Joules per mole of atmosphere (see results section). Because there are $1.8 \times 10^{20}$ moles in Earth's atmosphere, this implies the total available energy in Earth's atmosphere due to this pairing is $1.3 \times 1.8 \times 10^{20} = 2.34 \times 10^{20}$ J. The turnover lifetime of $CH_4$ in Earth's atmosphere is approximately 10 years (Dlugokencky *et al.* 1998; Prinn *et al.* 2001). If we assume that all the $CH_4$ in Earth's atmosphere is oxidized in 10 years ($3.15 \times 10^8$ seconds) on average, then the "power" from the $CH_4$-$O_2$ reaction according to our calculations will be the free energy release spread over this time: Power = $(2.34 \times 10^{20})/(3.15 \times 10^8) = 0.7$

TW. Simoncini *et al.* (2013) also found the power required to maintain the $O_2$-$CH_4$ disequilibrium to be 0.7 TW.

We have shown that large thermodynamic disequilibria coincides with surface biology in the Solar System, but whether chemical disequilibrium would be a useful metric for identifying exoplanet biospheres remains an open question. In principle, low-flux abiotic processes with slow kinetics could maintain a large atmospheric chemical disequilibrium. For example, the reaction between $H_2$ and $N_2$ is kinetically inhibited at Earth surface conditions, and so these species could coexist in thermodynamic disequilibrium for long timescales with very small replenishing fluxes (assuming a super Earth with sufficient gravity to retain hydrogen). In practice, however, there are few kinetic barriers to gas phase reactions at Earth-like temperatures and pressures, and so sizeable disequilibria from abiotic processes may be rare. In future work, it would be helpful to apply this metric to model exoplanet atmospheres to determine if there are any plausible false positives scenarios, i.e. dead worlds with large available Gibbs energy. For example, a Mars-like atmosphere with different CO, $O_2$ and $H_2$ abundances - perhaps due to elevated UV irradiation or different outgassed species - could perhaps have a large thermodynamic disequilibrium in the absence of life. Future work should also investigate how abiotic disequilibrium in Solar System atmospheres may have varied since 4.56 Ga.

*Practicality of thermodynamic disequilibrium as exoplanet biosignature*
The main advantage of using thermodynamic disequilibrium for biosphere detection over kinetic metrics is that it requires minimal auxiliary assumptions. Whereas kinetic arguments require abiotic surface sinks to be estimated, the calculation of gas phase chemical equilibrium in a planet's atmosphere requires only bulk atmosphere abundances, surface temperature, and pressure; future observations could be used to infer all three of these (Des Marais *et al.* 2002; Misra *et al.* 2014). Schwieterman *et al.* (2015) recently demonstrated that it is possible to constrain the abundance of molecular nitrogen due to its tendency to form $N_2$-$N_2$ dimers, which are spectrally active at 4.3 μm. Multiphase calculations for atmosphere-ocean systems require knowledge of a surface ocean. In principle, it is possible to infer the presence of exoplanet oceans using ocean glint (Robinson *et al.* 2010), and the approximate surface extent of oceans may be estimated with time-resolved photometry (Cowan *et al.* 2009). The sensitivity of our metric to ocean volume is discussed below. Recall also that our multiphase calculations do not fully capture the disequilibrium in the surface reservoirs since they neglect any reactions with the crust (see above).

In principle, the abundances of dissolved ions and ocean pH would also be required to calculate the atmosphere-ocean disequilibrium for an exo-Earth. However, the available Gibbs energy in Earth's atmosphere-ocean system is relatively insensitive to these variables. Table 10 shows the sensitivity of the available Gibbs energy in Earth's atmosphere-ocean system to variables that could not be measured remotely (or would be difficult to observe remotely) on exoplanets. Key findings are summarized below.

The available Gibbs energy of the Earth's atmosphere ocean system is largely insensitive to both ocean salinity and pH. Only at extremely low pH values (pH=2) does the available energy decrease by around 15% since the equilibrium of reaction (21) is pushed to the left. This insensitivity arises because the starting abundance of hydrogen ions is many orders of magnitude less than the equilibrium abundance, and so changes to the initial abundance (pH) does not affect the equilibrium state very much. This suggests the pH and salinity of exoplanet oceans do not need to be known to estimate the available Gibbs energy in their atmosphere-ocean systems.

The available Gibbs energy of the Earth system is moderately sensitive to ocean volume. Increasing ocean volume by a factor of 10 increases the available Gibbs energy of a factor of 4. The disequilibrium in an exoplanet atmosphere-ocean system could be overestimated if oceans are extremely shallow. For example, if the Earth's oceans were only 10% of their current volume then the available Gibbs energy in the Earth system would be 413 J/mol, only ~3 times larger than our value calculated for Mars (table 8).

Various observational techniques have been proposed to both detect oceans (Gaidos and Williams 2004; Robinson *et al.* 2010; Zugger *et al.* 2010), and to map the ocean-land fraction for terrestrial exoplanets using time resolved photometry (Cowan *et al.* 2009; Cowan and Strait 2013; Fujii and Kawahara 2012). These studies suggest that a ~10 m space telescope should be able to obtain a coarse surface map of an Earth-analog at 10 pc. Given observations of surface ocean fraction, it may be possible to constrain ocean depth from geophysical theory. For example, the typical strength or rock would not support a large topographic elevation between seafloor and land. In the case of granitic continents and a basaltic seafloor, the maximum possible ocean depth with exposed continents is approximately equal to $11.4 \times g_{Earth} / g_{planet}$ km, where $g_{Earth}$ is the surface gravity of the Earth and $g_{planet}$ is the surface gravity of the planet of interest (Cowan and Abbot 2014). Of course, for planets with no land and very deep oceans (~1000 km), ocean volume could be constrained by mass and radius observations. Putting a lower limit on ocean depth is more challenging, but several possibilities exist. Heat flow from planetary interiors is uneven due to the large spacing of convective cells in a viscous fluid, and will therefore inevitably create some topographic relief (Davies 1998). Consequently, the elevation distributions of the terrestrial Solar System planets all extend over several kilometers (Melosh 2011, p. 42). It may also be possible to put a lower bound on ocean volume using thermal inertia arguments and observed variations in a planet's infrared flux over its orbit (Gaidos and Williams 2004). By combining land-ocean maps, thermal inertia observations, and geophysical constraints on topography, estimates of ocean volume may be obtainable for some exoplanets, but solving this problem is beyond the scope of the current paper.

Our equilibrium calculation for the Earth atmosphere-ocean system is a simplification because we have assumed the entire atmosphere and ocean are at a mean global temperature and sea-level pressure (T=15°C and P=1.013 bar). In practice, the air temperature varies over the surface, ocean temperature typically decreases with depth, and pressure increases by several orders of magnitude in the deep ocean. To investigate the sensitivity of the available Gibbs energy to these variations we repeated the equilibrium calculation for a wide range of pressures and temperatures. The available Gibbs energy of the Earth's atmosphere-ocean system is moderately sensitive to these changes (table 10). At 0°C the available Gibbs energy is around 30% lower than the value at the observed mean surface temperature, 15°C. If temperature is instead increased to 25°C then the available Gibbs energy increases by around 20%. Changing the pressure by an order of magnitude in either direction results in a change in available Gibbs energy by approximately a factor of two, though at very high pressures (1000 bar) the available Gibbs energy asymptotes to a value of nearly 7000 J/mol. These results demonstrate that the available Gibbs energy of the Earth atmosphere-ocean system may be somewhat different if spatial variations in temperature and pressure are accounted for, but that our result (~2300 J/mol) is accurate to well within an order of magnitude. This sensitivity analysis also establishes that it is not necessary to determine the surface temperature and pressure of exoplanets with high precision to estimate the available Gibbs energy of their atmosphere-ocean systems.

In summary, sensitivity analysis suggests that with good observations it might be possible to calculate thermodynamic disequilibrium for exoplanets. The gas-phase calculations have no strong sensitivities to difficult-to-observe variables such as ocean volume, and so thermodynamic disequilibrium could be accurately calculated from remote observations. Additionally, gas phase reactions are much more weakly dependent on pressure and temperature than multiphase reactions. For multiphase calculations, it may be possible to estimate thermodynamic disequilibrium to the correct order of magnitude. An important caveat on this result is that the available Gibbs energy of the Earth is moderately sensitive to ocean volume, and it may be challenging to put a lower bound on ocean depth.

Future work will assess the sensitivity of our metric to potential uncertainties in the inferences from future telescopic observations, which are expected because of limitations such as spectroscopic resolution. Such work is beyond the scope of the current paper, the purpose of which is to describe our basic methodology and discuss results for Solar System bodies.

**Conclusions**
- We have quantified the atmospheric chemical disequilibrium for Solar System planets with thick atmospheres. The magnitude of the purely gas phase disequilibrium in Earth's atmosphere, 1.5 J/mol, is not unusual by Solar System standards.
- However, a multiphase equilibrium calculation reveals that the disequilibrium in Earth's atmosphere-ocean system, 2326 J/mol, is at least an order of magnitude larger than any other atmosphere in the Solar System. Note that we did not do a full multiphase calculation for Titan because the mean composition of all its hydrocarbon lakes is not known, so we are comparing the Gibbs energy of Earth's atmosphere-ocean system to other solar system atmospheres only.
- Earth's disequilibrium is not caused by $O_2$-$CH_4$ pairing (a contribution of only 1.3 J/mol) but rather by the disequilibrium between $N_2$-$O_2$-$H_2O(l)$. This disequilibrium is maintained by life. Oxygenic photosynthesis replenishes molecular oxygen and the oxidation of fixed nitrogen and biological denitrification prevents the accumulation of nitrate in the ocean.
- The atmospheric composition of terrestrial exoplanets will be accessible to future telescopes, and so gas phase thermodynamic disequilibrium may be readily calculated for these planets' atmospheres. It may also be possible to estimate the multiphase disequilibrium for exoplanets if surface oceans can be detected and volumetrically constrained.
- Further work will be required to evaluate the utility of thermodynamic disequilibrium as a generalized metric for surface biospheres.

**Appendices**

**Appendix A: Gas phase Gibbs minimization**
This section describes the methodology used to find gas phase equilibrium using Gibbs free energy minimization. We provide the Matlab code that implements this methodology on the website of the senior author (DCC). Recall that for a gas phase system, the equilibrium state has mole fraction abundances, $\bar{n}_i$, that minimize equation (5) in the main text. Temperature dependent standard Gibbs free energies of formation, $\Delta_f G^°_{i(T,P_r)}$, were calculated from enthalpies and entropies of formation retrieved from NASA's thermodynamic database (Burcat and Ruscic 2005). We used the 2009 version of this database (available here http://www.grc.nasa.gov/WWW/CEAWeb/ or http://garfield.chem.elte.hu/Burcat/NEWNASA.TXT). The database provides 10 coefficients for each gaseous species (sometimes multiple sets of 10 coefficients are specified for different temperature ranges). The enthalpies and entropies of formation are calculated from these coefficients using the following empirically fitted expressions:

$$\Delta_f H^°_{i(T,P_r)} / RT = -a_1 T^{-2} + a_2 \ln(T)/T + a_3 + a_4 T/2 + a_5 T^2/3 + a_6 T^3/4 + a_7 T^4/5 + a_9/T \quad (24)$$

$$\Delta_f S^°_{i(T,P_r)} / R = -a_1 T^{-2}/2 - a_2/T + a_3 \ln(T) + a_4 T + a_5 T^2/2 + a_6 T^3/3 + a_7 T^4/4 + a_{10} \quad (25)$$

Here, $a_{1-10}$ are the coefficients from the NASA database (the 9$^{th}$ coefficient is unused). Enthalpies and entropies are combined to calculate the Gibbs free energy of formation:

$$\Delta_f G^°_{i(T,P_r)} = \Delta_f H^°_{i(T,P_r)} - T \Delta_f S^°_{i(T,P_r)} \quad (26)$$

Note that there are several different conventions for Gibbs free energies of formation (see for instance Anderson and Crerar (1993, p154)). The different conventions produce equivalent equilibrium results, but it is important to use Gibbs energies of the same convention within any given calculation. The NASA database provides Gibbs free energies of formation according to the Berman-Brown convention (e.g. Anderson and Crerar (1993, p156)), but we convert these to standard free energies of formation in our Matlab code.

In the expression for Gibbs energy, equation (5), temperature dependent fugacity coefficients, $\gamma_{fi}$, were calculated using the Soave equation as described in Walas (1985, p146):

$$\ln(\gamma_f) = \frac{B_i}{B}(Z-1) - \ln(Z-B) + \frac{A}{B}\left[\frac{B_i}{B} - \frac{2}{a\alpha}\sum_j n_j (a\alpha)_{ij}\right] \ln\left(1 + \frac{B}{Z}\right) \quad (27)$$

Here, Z is the smallest real solution to the cubic, $f(Z) = Z^3 - Z^2 + (A - B - B^2)Z - AB = 0$. The other terms and variables are defined by the following set of equations:

$$\begin{aligned}
A &= (a\alpha)P/R^2T^2 & B &= bP/RT \\
B_i &= b_i P/RT & a\alpha &= \sum_i \sum_j n_i n_j (a\alpha)_{ij} \\
(a\alpha)_{ij} &= (1-k_{ij})\sqrt{(a_i\alpha_i)(a_j\alpha_j)} \\
a_i &= 0.42747\, R^2 T_{ci}^2 / P_{ci} \\
b_i &= 0.08664\, RT_{ci}/P_{ci} & b &= \sum_i n_i b_i \\
\alpha_i &= \left[1 + (0.480 + 1.574\omega_i - 0.176\omega_i^2)(1 - T_{ci}^{0.5})\right]^2
\end{aligned} \qquad (28)$$

In this set of equations, $n_i$ is the number of moles of the $i$-th species, $R$ is the universal gas constant, and $P$ and $T$ are the pressure and temperature of the system, respectively. $T_{ci}$ is the critical temperature of the $i$-th species and $P_{ci}$ is the critical pressure of the $i$-th species. Finally, $\omega_i$ is the acentric factor of the $i$-th species and $k_{ij}$ is a binary interaction parameter for species $i$ and $j$. All the other variables and terms are computable from these basic parameters. Critical temperatures, critical pressures, and acentric factors for gaseous species were obtained from Perry *et al.* (2008, section 2-136). To investigate the importance of binary interaction parameters, we performed some sensitivity tests using the simple gaseous system described in Lwin (2000). In this system, $H_2O$ and $CH_4$ are reacted to equilibrium to form $CO$, $CO_2$ and $H_2$ at high temperature (1000 K) and pressure (90 bar). We performed these tests at high temperature and pressure because this is the regime where departure from ideal behavior is the most significant. The inclusion of binary interaction parameters had a small effect on the fugacity coefficients and a negligible effect (<1%) on the overall change in Gibbs energy of the system. Consequently, in the equations above we assumed $k_{ij}=0$ for every pair of molecules. The close agreement between our numerical Gibbs free energy calculations, which don't include binary interaction parameter, and the Aspen Plus calculations, which do include binary interaction parameters, is further confirmation that ignoring binary interaction parameters is justified.

Given the Gibbs energies of formation and fugacity coefficients for all species, the Gibbs energy expression, equation (5), can be computed and minimized. We used an interior points method implemented using Matlab's *fmincon* function to minimize $\Delta G_{(T,P)}$. The $n_i$ that minimize $\Delta G_{(T,P)}$ and satisfy the atom conservation constraint (equation (6)) define the equilibrium state.

**Appendix B: Gibbs energy proof**

Here, we demonstrate that the minimum of equation (4) is identical to the minimum of equation (5). The standard Gibbs free energy of formation for a compound is the change in Gibbs energy with formation of one mole from its constituent elements in their standard states (i.e. the most stable elemental form at standard conditions, usually taken usually taken as 25°C and 1 atm for most databases (Anderson 2005, p. 211; Anderson and Crerar 1993, p. 154)):

$$\Delta_f G^\circ_{i\,(T,P_r)} \equiv G^\circ_{i\,(T,P_r)} - \sum_{elements=j} v_{ji} G^\circ_{j(T,P_r)} \qquad (29)$$

Recall that $v_{ji}$ is the number of atoms of element $j$ per molecule of species $i$, and $G^\circ_{j(T,P_r)}$ is the standard partial molar Gibbs free energy of gas $j$ at reference pressure $P_r$ and temperature $T$. The other variables are defined in the main text. Equation (29) can be substituted into equation (5) to obtain the following expression:

$$\begin{aligned}
\Delta G_{(T,P)} &= \sum_i n_i (\Delta_f G^\circ_{i\,(T,P_r)} + RT \ln(P_i \gamma_{fi})) \\
&= \sum_i n_i (G^\circ_{i\,(T,P_r)} - \sum_{elements=j} v_{ji} G^\circ_{j(T,P_r)} + RT \ln(P n_i \gamma_{fi} / n_T)) \\
&= G_{(T,P)} - \sum_i \sum_{elements=j} n_i v_{ji} G^\circ_{j(T,P_r)} = G_{(T,P)} - \sum_{elements=j} G^\circ_{j(T,P_r)} \sum_i n_i v_{ji}
\end{aligned} \qquad (30)$$

The last line uses equation (4) to substitute for $G_{(T,P)}$.

It is often assumed that $G^\circ_{j(T,P_r)} = 0$ for the elements, thereby establishing that $\Delta G_{(T,P)} = G_{(T,P)}$. However, this assumption is incorrect (Anderson and Crerar 1993, p. 147). Instead, the Gibbs free energy of formation for elements equals zero, $\Delta_f G^\circ_{i\,(T,P_r)} = 0$ if species $i$ is an element. Consequently, equations (4) and equation (5) are not identical $\left(\Delta G_{(T,P)} \neq G_{(T,P)}\right)$, but they do have the same minimum. This can be seen by considering the second term on the last line of equation (30). Conservation of atoms ensures that this term is a constant; refer to equation (6) if this is not immediately clear. Since this term is a constant, minimizing $G_{(T,P)}$ is equivalent to minimizing $\Delta G_{(T,P)}$. Note also that because $G_{(T,P)}$ and $\Delta G_{(T,P)}$ only differ by a constant, then differences in Gibbs energies between two states will be the same regardless of which form is used. This establishes that using equation (7) to calculate our metric of available Gibbs energies is equivalent to using equation (8).

**Appendix C: Multiphase calculations**

This section describes the methodology used to find multiphase equilibrium using Gibbs free energy minimization. Recall that for a multiphase system, the equilibrium state is the mole fraction abundances, $\bar{n}_i$, that minimize the expression in equation (9) in the main text. Temperature and pressure dependent standard Gibbs free energies of formation were calculated from the SUPCRT database (Johnson *et al.* 1992). Gibbs free energies of formation for aqueous species are given by the following expression (Walther 2009, p704):

$$\Delta_f G^\circ_{i\,(T,P)} = \Delta_f G^\circ_{i\,(T_r,P_r)} - \Delta_f S^\circ_{i\,(T_r,P_r)}(T-T_r) - c_1\left[T\ln\left(\frac{T}{T_r}\right)-T+T_r\right]$$

$$+ a_1(P-P_r) + a_2\ln\left(\frac{\Psi+P}{\Psi+P_r}\right)$$

$$- c_2\left\{\left[\left(\frac{1}{T-\theta}\right)-\left(\frac{1}{T_r-\theta}\right)\right]\left(\frac{\theta-T}{\theta}\right)-\frac{T}{\theta^2}\ln\left[\frac{T_r(T-\theta)}{T(T_r-\theta)}\right]\right\} \quad (31)$$

$$+ \left(\frac{1}{T-\theta}\right)\left[a_3(P-P_r) + a_4\ln\left(\frac{\Psi+P}{\Psi+P_r}\right)\right]$$

$$+ \omega_{P,T}\left(\frac{1}{\varepsilon_{P,T}}-1\right) - \omega_{P_r,T_r}\left(\frac{1}{\varepsilon_{P_r,T_r}}-1\right) + \omega_{P_r,T_r}Y_{P_r,T_r}(T-T_r)$$

Where:
$\Delta_f G^\circ_{i\,(T,P)}$ = Gibbs free energy of formation for the *i*-th species.
$\Delta_f G^\circ_{i\,(T_r,P_r)}$ = Gibbs free energy of formation for the *i*-th species at the reference temperature and pressure (from SUPCRT database)
$\Delta_f S^\circ_{i\,(T_r,P_r)}$ = Entropy of formation at the reference temperature and pressure
$T$ = Temperature of the system
$T_r$ = Reference temperature (298 K)
$P$ = Pressure of the system
$P_r$ = Reference pressure (1 bar)
$c_1, c_2, a_{1-4}$ = species specific coefficients (from the SUPCRT database)
$\Psi$ = Solvent pressure parameter (2600 bar)
$\theta$ = Solvent temperature parameter (228 K)
$\omega_{P,T}$ = Born coefficient
$\omega_{P_r,T_r}$ = Born coefficient at the reference temperature and pressure (from the SUPCRT database)
$\varepsilon_{P,T}$ = Dielectric constant of water.

$\varepsilon_{P_r,T_r}$ = Dielectric constant of water at the reference temperature and pressure (78.47)

$Y_{P_r,T_r}$ = Born derivative equation ($-5.81 \times 10^{-5}$ K$^{-1}$)

The Born coefficients have a small effect on the overall Gibbs energy of formation. For neutral species, $\omega_{P_r,T_r} = \omega_{P,T}$. In other cases, these two terms are nearly equal, and will approximately cancel each other. Thus, we simplified the Gibbs energy of formation expression by dropping these two terms:

$$\Delta_f G^\circ_{i\,(T,P)} = \Delta_f G^\circ_{i\,(T_r,P_r)} - \Delta_f S^\circ_{i\,(T_r,P_r)}(T-T_r) - c_1\left[T\ln\left(\frac{T}{T_r}\right) - T + T_r\right]$$
$$+ a_1(P-P_r) + a_2 \ln\left(\frac{\Psi + P}{\Psi + P_r}\right)$$
$$- c_2\left\{\left[\left(\frac{1}{T-\theta}\right) - \left(\frac{1}{T_r-\theta}\right)\right]\left(\frac{\theta-T}{\theta}\right) - \frac{T}{\theta^2}\ln\left[\frac{T_r(T-\theta)}{T(T_r-\theta)}\right]\right\} \quad (32)$$
$$+ \left(\frac{1}{T-\theta}\right)\left[a_3(P-P_r) + a_4 \ln\left(\frac{\Psi+P}{\Psi+P_r}\right)\right]$$
$$+ \omega_{P_r,T_r} Y_{P_r,T_r}(T-T_r)$$

This expression is used to calculate the Gibbs free energies of formation for all the aqueous species in our multiphase equilibrium code.

Activity coefficients for aqueous species, $\gamma_{ai}$ in equation (9), were approximated using the Truesdell-Jones equation (Langmuir 1997, p140):

$$\ln(\gamma_{ai}) = \frac{-0.5092 z_i^2 \sqrt{I}}{1 + 0.3283 a_i \sqrt{I}} + b_i I \quad (33)$$

Here, $I = \frac{1}{2}\sum_j m_j z_j^2$ is the ionic strength of the solution, where $m_j$ is the molality of the $j$-th species and $z_j$ is the charge of the $j$-th species. The variables $a_i$ and $b_i$ are species-specific thermodynamic coefficients that were obtained from Langmuir (1997). The Truesdell-Jones equation is only an approximation, but it is known to be accurate for solutions up to 2 molal (Langmuir 1997, p. 142). Because the Earth's ocean has an ionic strength of 0.7 molal, and the dissolution of nitrate and hydrogen by reaction to equilibrium does not increase this very much (see table 7), the Truesdell-Jones equation provides accurate activity coefficients in our calculations. Sensitivity analysis also reveals that the available

Gibbs free energy of the Earth is fairly insensitive to the activity coefficients of the major aqueous species. However, the available Gibbs energy of the Earth's atmosphere-ocean system is quite sensitive to water activity. Consequently, rather than use the Truesdell-Jones approximation above, the activity coefficient for water was calculated rigorously using a simplified form of the Pitzer equations (Marion and Kargel 2007):

$$\phi = 1 + \frac{2}{\sum_i m_i} \left\{ \frac{-0.3915 I^{3/2}}{1+1.2 I^{1/2}} + \sum_{all}\sum_{pairs} m_c m_a \left( B^\phi_{ca} + Z C_{ca} \right) \right\} \qquad (34)$$

Here, $\phi$ is the osmotic coefficient and can be related to the activity coefficient of water, $\gamma_{aw}$ in equation (9), by the following expression:

$$\ln(\gamma_{aw}) = -\phi \sum_i m_i / 55.50844 \qquad (35)$$

The double summation in equation (34) is over all unique pairs of anions and cations in solution (no double counting). The other variables in equation (34) are defined as follows:
$m_i$ = molality of the $i$-th species
$m_a$ = molality of the anion
$m_c$ = molality of the cation
$I$ = ionic strength of the solution (defined above)
$Z = \sum_i m_i |z_i|$
$B_{MX} = B^{(0)}_{MX} + B^{(1)}_{MX} \exp\left(-\alpha_1 I^{1/2}\right) + B^{(2)}_{MX} \exp\left(-\alpha_2 I^{1/2}\right)$
$\alpha_1 = 2.0\, kg^{0.5} mol^{-0.5}, \alpha_2 = 0\, kg^{0.5} mol^{-0.5}$ for all binary systems except 2:2 electrolytes
$\alpha_1 = 1.4\, kg^{0.5} mol^{-0.5}, \alpha_2 = 12\, kg^{0.5} mol^{-0.5}$ for 2:2 electrolytes
$B^{(0)}_{MX}, B^{(1)}_{MX}, B^{(2)}_{MX}, C_{MX}$ are species-specific binary interaction parameters that were obtained from Appelo and Postma (2005) and Marion (2002). The form of the Pitzer equation described above is a simplification of the complete expression; we have ignored cation-cation and anion-anion interactions, neutral solute parameters, and triple particle parameters since these terms will be small for Earth's ocean. Temperature dependencies were also ignored since, in absolute Kelvin, the temperature of the ocean is close to the reference temperature of 298 K. The activity coefficient of water was calculated using these equations at every iteration in our multiphase Gibbs free energy minimization calculations.

Finding the equilibrium state for multiphase systems is more challenging than for single-phase gaseous systems. The Matlab function *fmincon* was once again used to implement the optimization, but this time we provided the analytic gradient for the Gibbs energy function in equation (9) from differentiation, as follows:

$$\frac{1}{RT}\frac{\partial \Delta G_{(T,P)}}{\partial n_i} = \begin{cases} \frac{\Delta_f G^\circ_{i\,(T,P)}}{RT} + \ln(\gamma_{aw}) + \ln\left(\frac{n_i}{n_{aq}}\right) - \frac{n_i}{n_{aq}} - \frac{n_{aq}}{n_i} + 2 & \alpha = water \\ \frac{\Delta_f G^\circ_{i\,(T,P)}}{RT} + \ln\left(\frac{n_i}{n_\alpha}\right) + \ln(\gamma_{fi}) & \alpha = gas \\ \frac{\Delta_f G^\circ_{i\,(T,P)}}{RT} + \ln(55.5084) + \ln(\gamma_{ai}) + \ln\left(\frac{n_i}{n_{aq}}\right) - \ln\left(\frac{n_w}{n_{aq}}\right) - \frac{n_w}{n_{aq}} + 1 & \alpha = aqueous \end{cases}$$

(36)

The terms in this expression are defined in the methods section. We assumed that the activity coefficients were, to first order, independent of molar abundances.

Proving *fmincon* with an analytic gradient ensured more rapid and reliable convergence. For multiphase Gibbs energy minimization problems there is no guarantee that the local minima equal the global minimum (Nichita *et al.* 2002). Consequently, we implemented a simple global minimum search by iterating over an ensemble of random initial conditions. The vast majority of runs converged to the same minimum; only occasionally would an optimization run converge to another, less optimum, minimum or simply not converge. This gives us confidence that the consensus minimum was in fact the true global minimum. Semi-analytic calculations and Aspen Plus results also validate our multiphase Gibbs energy minimization result.

**Appendix D: Semi-analytic calculations**
Here, we describe the methodology for our semi-analytic calculation using equilibrium reactions in the atmospheres of Jupiter and Earth as examples. The reactions chosen for the semi-analytic calculations for the other atmospheres are also listed at the end.

*Jupiter*
In Jupiter's atmosphere, the key available redox couples suggest that there are two important reactions that contribute to chemical disequilibrium:

$$3H_2(g) + HCN(g) \rightleftharpoons CH_4(g) + NH_3(g) \quad (37)$$

$$3H_2 + CO \rightleftharpoons CH_4 + H_2O \quad (38)$$

We begin with reaction (37). The Gibbs energy of this reaction is given by (e.g. Anderson and Crerar (1993, p238)):

$$\Delta_r G = \Delta_r G° + RT \ln(Q) = \Delta_r G° + RT \ln\left(\frac{a_{CH_4} a_{NH_3}}{a_{H_2}^3 a_{HCN}}\right) \quad (39)$$

The activity of each species $i$ is denoted by $a_i$, the temperature of Jupiter's atmosphere at 1 bar is $T=165$ K, $R$ is the universal gas constant, and $Q$ is the reaction quotient. From equation (37) the Gibbs energy of the reaction, $\Delta_r G$, is the change in Gibbs energy of the system per 3 moles of $H_2$ and 1 mole of HCN that are converted to $CH_4$ and $NH_3$. The standard free energy of the reaction, $\Delta_r G°$, represents the value of this quantity when the activities of all species equals unity. In this case, taking $T = 165$ K and $P_r = 1$ bar,

$$\begin{aligned}\Delta_r G° &= \Delta_f G°_{CH_4(T,P_r)} + \Delta_f G°_{NH_3(T,P_r)} - 3\Delta_f G°_{H_2(T,P_r)} - \Delta_f G°_{HCN(T,P_r)} \\ &= -6.025 \times 10^4 + -2.88021 \times 10^4 - 3 \times 0 - 1.27374866 \times 10^5 \text{ J/mol} \\ &= -2.1643 \times 10^5 \text{ J/mol}\end{aligned} \quad (40)$$

where we have substituted the appropriate Gibbs free energies of formation for each species computed at 165 K using the database and methodology of Appendix A. Gibbs free energies of formation were taken from the same thermodynamic databases as those used for the Gibbs energy minimization calculations.

Reaction (37) is in equilibrium when the left hand side of equation (39) is zero. We solve for this equilibrium by making the following substitution:

$$\Delta_r G(x) = \Delta_r G° + RT \ln\left(\frac{\left[P(n_{CH_4} + x)/n_T\right]\left[P(n_{NH_3} + x)/n_T\right]}{\left[P(n_{H_2} - 3x)/n_T\right]^3 \left[P(n_{HCN} - x)/n_T\right]}\right) \quad (41)$$

Here, $n_i$ is the observed moles for each species, $n_T$ is the total number of moles, $P$ is the pressure, and $x$ is the number of moles that have reacted. We solve for $x$ to find the equilibrium abundances for each species. Note that since we are performing this calculation at $P=1$ bar in Jupiter's atmosphere, and since we are using mixing ratios for the number of moles ($n_T = 1$) the equation (41) can be simplified:

$$\Delta_r G(x) = \Delta_r G° + RT \ln\left(\frac{(n_{CH_4} + x)(n_{NH_3} + x)}{(n_{H_2} - 3x)^3 (n_{HCN} - x)}\right) \quad (42)$$

By setting $\Delta_r G(x) = 0$ this equation can be rearranged to give the following polynomial:

$$\left(n_{H_2} - 3x\right)^3 \left(n_{HCN} - x\right)\exp\left(-\frac{\Delta_r G^\circ}{RT}\right) - \left(n_{CH_4} + x\right)\left(n_{NH_3} + x\right) = 0 \quad (43)$$

This polynomial in $x$ is solved numerically. The equilibrium is the smallest real solution since the reaction will proceed to this point. In this case this solution is $x_{eqm}=3.6\times10^{-9}$. This solution equals the initial mixing ratio of HCN (table 3), which implies that reaction (37) goes to completion when Jupiter's atmosphere is reacted to equilibrium.

To calculate the change in Gibbs energy change associated with this reaction going to completion we calculate the integral:

$$\int_{x=0}^{x=3.6\times10^{-9}} \Delta_r G(x')/n_T dx' = \int_{x=0}^{x=3.6\times10^{-9}} \Delta_r G(x')dx'$$

$$= \int_{x=0}^{x=3.6\times10^{-9}} \left[\Delta_r G^\circ + RT \ln\left(\frac{\left(n_{CH_4} + x'\right)\left(n_{NH_3} + x'\right)}{\left(n_{H_2} - 3x'\right)^3 \left(n_{HCN} - x'\right)}\right)\right] dx' = 7.5137\times10^{-4} \; J/mol$$
(44)

The same methodology can be repeated for reaction (38).

$$\Delta_r G = \Delta_r G^\circ + RT\ln(Q) = \Delta_r G^\circ + RT\ln\left(\frac{a_{CH_4} a_{H_2O}}{a_{H_2}^3 a_{CO}}\right) \quad (45)$$

In this case the standard free energy of the reaction computed at $T = 165$ K is $\Delta_r G^\circ = -1.68862\times10^5$ J/mol. Substituting activities for x and simplifying yields the equation:

$$\Delta_r G(x) = \Delta_r G^\circ + RT\ln\left(\frac{\left(n_{CH_4} + x\right)\left(n_{H_2O} + x\right)}{\left(n_{H_2} - 3x\right)^3 \left(n_{CO} - x\right)}\right) \quad (46)$$

Next, the Gibbs energy of the reaction is set to zero and terms are rearranged to obtain the polynomial:

$$\left(n_{H_2} - 3x\right)^3 \left(n_{CO} - x\right)\exp\left(-\frac{\Delta_r G^\circ}{RT}\right) - \left(n_{CH_4} + x\right)\left(n_{H_2O} + x\right) = 0 \quad (47)$$

The solution to this polynomial is $x_{eqm}=1.6\times10^{-9}$ which indicates that CO is depleted and this reaction also goes to completion. The change in Gibbs free energy associated with this reaction is given by:

$$\int_{x=0}^{x=1.6\times10^{-9}} \Delta_r G(x)dx = 2.8068\text{e-}04\times10^{-4} \; J/mol \quad (48)$$

Finally, we sum together the Gibbs energy changes from these two reactions to obtain an approximation of the available Gibbs energy in Jupiter's atmosphere:

$$\Phi \approx 7.5137 \times 10^{-4} + 2.8068\text{e-}04 \times 10^{-4} = 0.001032 \text{ J/mol} \quad (49)$$

This compares to 0.001032 J/mol using the numerical model (main text, table 3), so the semi-analytic approximation is good to 4 significant figures in this instance.

*Earth (atmosphere-ocean):*

Next, we describe our semi-analytic calculations for the Earth atmosphere-ocean system. These calculations were used to obtain the "semi-analytic approximation" values in table 8 and table 9. Firstly, we consider the Gibbs energy associated with the equation:

$$2\text{N}_2(\text{g}) + 5\text{O}_2(\text{g}) + 2\text{H}_2\text{O}(\text{l}) \rightleftharpoons 4\text{H}^+(\text{aq}) + 4\text{NO}_3^-(\text{aq}) \quad (50)$$

The Gibbs energy of this reaction is given by:

$$\Delta_r G = \Delta_r G^\circ + RT \ln(Q) = \Delta_r G^\circ + RT \ln\left(\frac{a_{H^+}^4 a_{NO_3^-}^4}{a_{N_2}^2 a_{O_2}^5 a_{H_2O}^2}\right) \quad (51)$$

The activity of each species $i$ is denoted by $a_i$, the average temperature of Earth's atmosphere at the surface is $T=288.15$ K, $R$ is the universal gas constant, and $Q$ is the reaction quotient. From equation (50) the Gibbs energy of the reaction, $\Delta_r G$, is the change in Gibbs energy of the system per 2 moles of N$_2$, 5 moles of O$_2$, and 2 moles of H$_2$O$_{(l)}$ that are converted to hydrogen ions and nitrate. The standard free energy of the reaction, $\Delta_r G^\circ$, represents the value of this quantity when the activities of all species equals unity. In this case, with $T = 288.15$ K and $P = P_r = 1$ bar:

$$\Delta_r G^\circ = 4\Delta_f G^\circ_{H^+(T,P)} + 4\Delta_f G^\circ_{NO_3^-(T,P)} - 2\Delta_f G^\circ_{H_2O_{(l)}(T,P_r)} - 5\Delta_f G^\circ_{O_2(T,P_r)} - 2\Delta_f G^\circ_{N_2(T,P_r)}$$

$$= 4 \times 0 + 4 \times -1.09164 \times 10^5 - 2 \times -2.387764 \times 10^5 - 5 \times 0 - 2 \times 0 \text{ J/mol}$$

$$= 4.0897 \times 10^4 \text{ J/mol}$$

(52)

Gibbs free energies of formation were taken from the same thermodynamic databases as those used for the Gibbs energy minimization calculations. Reaction (50) is in equilibrium when the left hand side of equation (51) is equal to zero. We solve for this equilibrium by making the following substitution:

$$\Delta_r G(x) = \Delta_r G^\circ + RT \ln\left(\frac{\left[\gamma_{H^+}\left(n_{H^+} + 4x\right)/M_{ocean}\right]^4 \left[\gamma_{NO_3^-}\left(n_{NO_3^-} + 4x\right)/M_{ocean}\right]^4}{\left[P(n_{N_2} - 2x)/n_T\right]^2 \left[P(n_{O_2} - 5x)/n_T\right]^5 a_{H_2O}^2}\right)$$

(53)

The activities of aqueous species are given by their molalities multiplied by an activity coefficient. Here, $M_{ocean} = 1.3802 \times 10^{21}$ kg is the mass of the Earth's ocean, $n_i$ is the observed moles for each species, $n_T = 1.7560 \times 10^{20}$ is the total number of moles of air (all gases) in the atmosphere, $P=1.013$ bar is the mean sea-level pressure, and x is the number of moles that have reacted. We solve for $x$ to find the equilibrium abundances for each species. By setting the left hand side of equation (53) to zero, assuming that the activity of water equals 1, and that the activity coefficients of all other species are 1, we obtain the following polynomial in $x$:

$$\left(\frac{P}{n_T}\right)^7 (n_{N_2} - 2x)^2 (n_{O_2} - 5x)^5 e^{-\frac{\Delta_r G^\circ}{RT}} = \left(\frac{1}{M_{ocean}^8}\right)(n_{H^+} + 4x)^4 (n_{NO_3} + 4x)^4 \tag{54}$$

This polynomial is solved numerically. The equilibrium is the smallest real solution since the reaction will proceed to this point. In this case this solution is $x_{eqm} = 6.05586 \times 10^{18}$. This solution does not equal the initial mixing ratio of $O_2$, which implies that reaction does not go all the way to completion.

To calculate the change in Gibbs energy change associated with reaction (50) going to equilibrium we calculate the integral:

$$\Delta G_1 = \int_{x=0}^{x=6.055 \times 10^{18}} \Delta_r G(x)/n_T \, dx = 1051 \text{ J/mol} \tag{55}$$

This is how the "semi-analytic approximation" value in row 1, table 9 was calculated.

Next, we consider the Gibbs energy changes associated with the following carbon-bearing reactions:

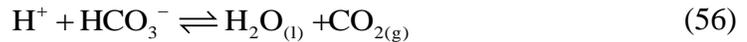
$$H^+ + HCO_3^- \rightleftharpoons H_2O_{(l)} + CO_{2(g)} \tag{56}$$
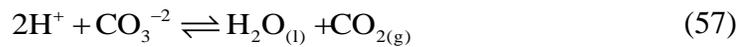
$$2H^+ + CO_3^{-2} \rightleftharpoons H_2O_{(l)} + CO_{2(g)} \tag{57}$$

The Gibbs energy of reaction (56), the first dissociation of carbonic acid, is given by:

$$\Delta_r G = \Delta_r G^\circ + RT \ln(Q) = \Delta_r G^\circ + RT \ln\left(\frac{a_{H_2O_{(l)}} a_{CO_{2(g)}}}{a_{H^+} a_{HCO_3^-}}\right) \tag{58}$$

The method for calculating the Gibbs energy change for this reaction is identical to that described above, and so we simply list the key equations:

$$\Delta_r G^\circ = \Delta_f G^\circ_{H_2O_{(l)}(T,P_r)} + \Delta_f G^\circ_{CO_{2(g)}(T,P_r)} - \Delta_f G^\circ_{H^+(T,P)} - 5\Delta_f G^\circ_{HCO_3^-(T,P)}$$
$$= 4.7475 \times 10^4 \text{ J/mol, with } T = 288 \text{ K and } P = P_r = 1 \text{ bar} \tag{59}$$

$$0 = \Delta_r G(x) = \Delta_r G° + RT \ln\left(\frac{\left[P(n_{CO_2}+x)/n_T\right]a_{H_2O}}{\left[\gamma_{H^+}(n_{H^+}-x)/M_{ocean}\right]\left[\gamma_{HCO_3^-}(n_{HCO_3^-}-x)/M_{ocean}\right]}\right) \quad (60)$$

Crucially, $n_{H^+}$ is not the observed H$^+$ abundance but is instead the equilibrium abundance from reaction (50); it is the acidification of the ocean from dissolved nitrate that drives the change in carbon species (see main text). Simplifying to obtain polynomial in $x$:

$$(n_{H^+}-x)(n_{HCO_3^-}-x)\exp\left(\frac{-\Delta_r G°}{RT}\right) = M_{ocean}^2 P(n_{CO_2}+x)/n_T \quad (61)$$

The physically relevant solution is $x_{eqm}= 2.3951\times 10^{18}$. The Gibbs energy change for the reaction can thus be calculated:

$$\Delta G_2 = \int_{x=0}^{x=2.44\times 10^{18}} \Delta_r G(x)/n_T \, dx = 520 \text{ J/mol} \quad (62)$$

Repeating this procedure for equation (57), the second dissociation of carbonic acid, yields $\Delta G_3 = 152 \text{ J/mol}$. The contributions from all three reactions can be summed to approximate the total available Gibbs energy for the Earth atmosphere-ocean system (assuming water activity equals 1):

$$\Phi \approx \Delta G_1 + \Delta G_2 + \Delta G_3 = 1724 \text{ J/mol} \quad (63)$$

This is how the value for the "semi-analytic approximation" in tables 8 and 9 was calculated. There is no straightforward way to extend this semi-analytic method to include changes in water activity, hence the discrepancy between semi-analytic and numerical values.

This procedure was repeated to approximate the available Gibbs energy for the Solar System planets. The key redox reactions chosen for these calculations are as follows:

*Mars:*

$$2CO + O_2 \rightleftharpoons 2CO_2 \quad (64)$$
$$2NO \rightleftharpoons N_2 + O_2 \quad (65)$$
$$2O_3 \rightleftharpoons 3O_2 \quad (66)$$
$$2H_2 + O_2 \rightleftharpoons 2H_2O \quad (67)$$

In Mars' case all these reactions go to completion.

*Venus:*

$$H_2S + CO_2 \rightleftharpoons H_2O + OCS \quad (68)$$
$$3CO + SO_2 \rightleftharpoons 2CO_2 + OCS \quad (69)$$

$$2CO_2 + S \rightleftharpoons SO_2 + 2CO \tag{70}$$

$$H_2 + S \rightleftharpoons H_2S \tag{71}$$

In Venus' case, reactions (70) and (71) go to completion whereas reactions (68) and (69) reach equilibria where the reactants are not entirely depleted.

*Earth (atmosphere only):*

$$CH_4 + 2O_2 \rightleftharpoons 2H_2O + CO_2 \tag{72}$$

$$2H_2 + O_2 \rightleftharpoons 2H_2O \tag{73}$$

$$2CO + O_2 \rightleftharpoons 2CO_2 \tag{74}$$

$$2N_2O \rightleftharpoons 2N_2 + O_2 \tag{75}$$

All of these reactions proceed to completion in the Earth's case.

*Titan:*

$$C_2H_6 + H_2 \rightleftharpoons 2CH_4 \tag{76}$$

$$C_2H_2 + 3H_2 \rightleftharpoons 2CH_4 \tag{77}$$

Both of these reactions proceed to completion.

*Uranus:*

$$C_2H_6 + H_2 \rightleftharpoons 2CH_4 \tag{78}$$

$$C_2H_2 + 3H_2 \rightleftharpoons 2CH_4 \tag{79}$$

$$CO + 3H_2 \rightleftharpoons CH_4 + H_2O \tag{80}$$

All three reactions proceed to completion.

**Appendix E: Multiphase calculations in Aspen Plus**

To validate our multiphase Matlab calculations, we used Aspen Plus to calculate chemical and phase equilibrium for the Earth atmosphere-ocean system. Figure E1 shows the Aspen Plus flowsheet. The observed state was partitioned into vapor and liquid phases, and fed into the RGIBBS reactor as two separate streams. RGIBBS is a module in Aspen Plus that can calculate equilibrium abundances using Gibbs free energy minimization. The resultant mixed stream was fed into a Flash2 phase separator and partitioned into equilibrium vapor abundances and liquid abundances. Without the phase separator the equilibrium results were unphysical, and the resultant Gibbs energy change was inaccurate. We used a calculator block to determine the Gibbs energy change between the two input streams and two output streams. Calculator blocks were necessary to compute the Gibbs energy of the initial and equilibrium states with sufficient precision to calculate the Gibbs energy change accurately (otherwise the default output did not provide enough significant figures).

To check that our results were robust we used the setup of figure E1 to calculate the equilibrium state using two different Aspen Plus electrolyte models, the Electrolyte Non-Random Two Liquid (ELECNRTL) model the PITZER model. Henry's Law components were used for all gaseous species except water. The equilibrium abundances from both models were very similar. The overall Gibbs energy change of the Earth atmosphere-ocean system was 2348 J/mol for the ELECNRTL model and 2205 J/mol for the PITZER model. It is unsurprising that there are slight differences between the two models since they use different equations of state and different thermodynamic property models. Both agree with our own numerical Gibbs energy minimization to within 6%.

**Acknowledgements**
We thank Jonathan D. Toner for assistance with aqueous thermodynamics and implementing the Pitzer equations. We also thank Kathryn Cogert for assistance with Aspen Plus, and the two anonymous reviewers, whose comments greatly improved the clarity and science content of this manuscript. DCC thanks Chris Glein for some assistance on preliminary work at the inception of this project in 2005. This work was supported by Exobiology Program grant NNX10AQ90G awarded to DCC and the NASA Astrobiology Institute's Virtual Planetary Laboratory under Cooperative Agreement Number NNA13AA93A. DSB was also supported by the NASA Washington Space Grant when he was an undergraduate researcher with DCC at the University of Washington.

**Disclosure statement**
No competing financial interests exist.

**References**
Anderson GM. (2005) Thermodynamics of natural systems. Cambridge University Press.
Anderson GM, and Crerar DA. (1993) Thermodynamics in geochemistry: the equilibrium model. Oxford University Press.
Appelo CAJ, and Postma D. (2005) Geochemistry, groundwater and pollution. CRC press.
Aspen Technology Inc. D. (2000) Aspen Plus: Getting Started Modeling Processes with Electrolytes. Aspen Technology, Inc., Cambridge, MA.
Bada JL, and Miller SL. (1968) Ammonium ion concentration in the primitive ocean. *Science*, 159, 423-425.
Baines KH, Atreya SK, Bullock MA, Grinspoon DH, Mahaffy P, Russell CT, Schubert G, and Zahnle K. (2014) The Atmospheres of the Terrestrial Planets: Clues to the Origins and Early Evolution of Venus, Earth, and Mars. *Comparative Climatology of Terrestrial Planets*, 137.

Balzhiser RE, Samuels MR, and Eliassen JD. (1972) Chemical engineering thermodynamics; the study of energy, entropy, and equilibrium. Prentice-Hall, Englewood Cliffs, N.J.,.
Barman T. (2007) Identification of absorption features in an extrasolar planet atmosphere. *The Astrophysical Journal Letters*, 661, L191.
Belu A, Selsis F, Morales J-C, Ribas I, Cossou C, and Rauer H. (2011) Primary and secondary eclipse spectroscopy with JWST: exploring the exoplanet parameter space. *Astronomy & Astrophysics*, 525, A83.
Bézard B, Lellouch E, Strobel D, Maillard J-P, and Drossart P. (2002) Carbon monoxide on Jupiter: Evidence for both internal and external sources. *Icarus*, 159, 95-111.
Bougher SW, Hunten DM, and Phillips RJ. (1997) Venus II--geology, Geophysics, Atmosphere, and Solar Wind Environment. University of Arizona Press.
Burcat A, and Ruscic B. (2005) Third Millennium ideal gas and condensed phase thermochemical database for combustion with updates from active thermochemical tables. US Dept. of Energy, Oak Ridge, TN.
Byrd RH, Gilbert JC, and Nocedal J. (2000) A trust region method based on interior point techniques for nonlinear programming. *Mathematical Programming*, 89, 149-185.
Byrd RH, Hribar ME, and Nocedal J. (1999) An interior point algorithm for large-scale nonlinear programming. *SIAM Journal on Optimization*, 9, 877-900.
Capone DG, Popa R, Flood B, and Nealson KH. (2006) Geochemistry. Follow the nitrogen. *Science (New York, NY)*, 312, 708.
Catling D, and Bergsman D. (2009) Using atmospheric composition as a metric for detecting life on habitable planets.AGU Fall Meeting Abstracts.
Catling D, and Bergsman D. (2010) On detecting exoplanet biospheres from atmospheric chemical disequilibrium. *LPI Contributions*, 1538, 5533.
Catling D, and Kasting JF. (2007) Planetary atmospheres and life. In: *Planets and Life: The Emerging Science of Astrobiology*. edited by WT Sullivan and JA Barosss, Cambridge Univ. Press, Cambridge, p 91-116.
Catling DC. (2015) Planetary Atmospheres. In: *Treatise on Geophysics*. edited by G Schuberts, Elsevier, in press.
Catling DC, Zahnle KJ, and McKay CP. (2001) Biogenic methane, hydrogen escape, and the irreversible oxidation of early Earth. *Science*, 293, 839-843.
Charbonneau D, Brown TM, Noyes RW, and Gilliland RL. (2002) Detection of an extrasolar planet atmosphere. *The Astrophysical Journal*, 568, 377.
Cockell CS, Léger A, Fridlund M, Herbst T, Kaltenegger L, Absil O, Beichman C, Benz W, Blanc M, and Brack A. (2009) Darwin-a mission to detect and search for life on extrasolar planets. *Astrobiology*, 9, 1-22.

Cordier D, Mousis O, Lunine JI, Lavvas P, and Vuitton V. (2009) An estimate of the chemical composition of Titan's lakes. *The Astrophysical Journal Letters*, 707, L128.

Covey C, Haberle RM, McKay CP, and Titov DV. (2013) The greenhouse effect and climate feedbacks. In: *Comparative Climatology of Terrestrial Planets*. edited by SJ Mackwell, AA Simon-Miller, JW Harder and MA Bullocks, University of Arizona Press.

Cowan NB, and Abbot DS. (2014) Water Cycling Between Ocean and Mantle: Super-Earths Need Not be Waterworlds. *arXiv preprint arXiv:1401.0720*.

Cowan NB, Agol E, Meadows VS, Robinson T, Livengood TA, Deming D, Lisse CM, A'Hearn MF, Wellnitz DD, and Seager S. (2009) Alien maps of an ocean-bearing world. *The Astrophysical Journal*, 700, 915.

Cowan NB, and Strait TE. (2013) Determining reflectance spectra of surfaces and clouds on exoplanets. *The Astrophysical Journal Letters*, 765, L17.

Davies GF. (1998) Topography: a robust constraint on mantle fluxes. *Chemical geology*, 145, 479-489.

Deming D, Seager S, Winn J, Miller-Ricci E, Clampin M, Lindler D, Greene T, Charbonneau D, Laughlin G, and Ricker G. (2009) Discovery and characterization of transiting super Earths using an all-sky transit survey and follow-up by the James Webb Space Telescope. *Publications of the Astronomical Society of the Pacific*, 121, 952-967.

Deming D, Wilkins A, McCullough P, Burrows A, Fortney JJ, Agol E, Dobbs-Dixon I, Madhusudhan N, Crouzet N, and Desert J-M. (2013) Infrared transmission spectroscopy of the exoplanets HD 209458b and XO-1b using the Wide Field Camera-3 on the Hubble Space Telescope. *The Astrophysical Journal*, 774, 95.

Des Marais DJ, Harwit MO, Jucks KW, Kasting JF, Lin DN, Lunine JI, Schneider J, Seager S, Traub WA, and Woolf NJ. (2002) Remote sensing of planetary properties and biosignatures on extrasolar terrestrial planets. *Astrobiology*, 2, 153-181.

Devol AH. (2008) Nitrogen in the marine environment. In: *Nitrogen in the marine environment*. edited by DG Capone, DA Bronk, MR Mulholland and EJ Carpenters, Academic Press, p 263-301.

Dlugokencky E, Masarie K, Lang P, and Tans P. (1998) Continuing decline in the growth rate of the atmospheric methane burden. *Nature*, 393, 447-450.

Eriksson G. (1971) Thermodynamics studies of high temperature equilibria. 3. SOLGAS, a computer program for calculating composition and heat condition of an equilibrium mixture. *Acta chemica scandinavica*, 25, 2651-&.

Eriksson G. (1975) Thermodynamic studies of high-temperature equilibria. 12. SOLGASMIX, a computer-program for calculation of equilibrium compositions in multiphase systems. *Chemica Scripta*, 8, 100-103.

Estrada E. (2012) Returnability as a criterion of disequilibrium in atmospheric reactions networks. *Journal of Mathematical Chemistry*, 50, 1363-1372.

Fegley B. (2014) Venus. In: *Treatise on Geochemistry (Second Edition)*. edited by H Holland and KK Turekians, p 127-148.

Fraine J, Deming D, Benneke B, Knutson H, Jordán A, Espinoza N, Madhusudhan N, Wilkins A, and Todorov K. (2014) Water vapour absorption in the clear atmosphere of a Neptune-sized exoplanet. *Nature*, 513, 526-529.

Fujii Y, and Kawahara H. (2012) Mapping Earth analogs from photometric variability: Spin-orbit tomography for planets in inclined orbits. *The Astrophysical Journal*, 755, 101.

Gaidos E, and Williams D. (2004) Seasonality on terrestrial extrasolar planets: inferring obliquity and surface conditions from infrared light curves. *New Astronomy*, 10, 67-77.

Glein CR, and Shock EL. (2013) A geochemical model of non-ideal solutions in the methane–ethane–propane–nitrogen–acetylene system on Titan. *Geochimica et Cosmochimica Acta*, 115, 217-240.

Gruber N. (2008) The marine nitrogen cycle: overview and challenges. *Nitrogen in the marine environment*, 1-50.

Hayes JM, and Waldbauer JR. (2006) The carbon cycle and associated redox processes through time. *Philosophical Transactions of the Royal Society B: Biological Sciences*, 361, 931-950.

Hedelt P, von Paris P, Godolt M, Gebauer S, Grenfell JL, Rauer H, Schreier F, Selsis F, and Trautmann T. (2013) Spectral features of Earth-like planets and their detectability at different orbital distances around F, G, and K-type stars. *Astronomy & Astrophysics*, 553, A9.

Hitchcock DR, and Lovelock JE. (1967) Life detection by atmospheric analysis. *Icarus*, 7, 149-159.

Huguenin RL, Prinn RG, and Maderazzo M. (1977) Mars: photodesorption from mineral surfaces and its effects on atmospheric stability. *Icarus*, 32, 270-298.

Hutchinson G. (1954) The biochemistry of the terrestrial atmosphere. In: *The earth as a planet*. edited by GP Kuipers, The University of Chicago Press, Chicago, p 371.

Irwin P. (2009) Giant Planets of Our Solar System: Atmospheres, Composition, and Structure. Springer Science & Business Media.

Jacob D. (1999) Introduction to atmospheric chemistry. Princeton University Press.

Johnson JW, Oelkers EH, and Helgeson HC. (1992) SUPCRT92: A software package for calculating the standard molal thermodynamic properties of minerals, gases, aqueous species, and reactions from 1 to 5000 bar and 0 to 1000 C. *Computers & Geosciences*, 18, 899-947.

Karpov IK, Chudnenko KV, and Kulik DA. (1997) Modeling chemical mass transfer in geochemical processes; thermodynamic relations, conditions of equilibria and numerical algorithms. *American Journal of Science*, 297, 767-806.

Kasting J, Traub W, Roberge A, Leger A, Schwartz A, Wootten A, Vosteen A, Lo A, Brack A, and Tanner A. (2009) Exoplanet Characterization and the Search for Life.Astro2010: The Astronomy and Astrophysics Decadal Survey.

Kasting JF, Whitmire DP, and Reynolds RT. (1993) Habitable zones around main sequence stars. *Icarus*, 101, 108-128.

Kaye JA, and Strobel DF. (1983) HCN formation on Jupiter: The coupled photochemistry of ammonia and acetylene. *Icarus*, 54, 417-433.

Kirschke S, Bousquet P, Ciais P, Saunois M, Canadell JG, Dlugokencky EJ, Bergamaschi P, Bergmann D, Blake DR, and Bruhwiler L. (2013) Three decades of global methane sources and sinks. *Nature Geoscience*, 6, 813-823.

Kleidon A. (2012) How does the Earth system generate and maintain thermodynamic disequilibrium and what does it imply for the future of the planet? *Philosophical Transactions of the Royal Society A: Mathematical, Physical and Engineering Sciences*, 370, 1012-1040.

Knutson HA, Lewis N, Fortney JJ, Burrows A, Showman AP, Cowan NB, Agol E, Aigrain S, Charbonneau D, and Deming D. (2012) 3.6 and 4.5 μm phase curves and evidence for non-equilibrium chemistry in the atmosphere of extrasolar planet HD 189733b. *The Astrophysical Journal*, 754, 22.

Krasnopolsky VA. (2015) Vertical profiles of H 2 O, H 2 SO 4, and sulfuric acid concentration at 45–75km on Venus. *Icarus*, 252, 327-333.

Krasnopolsky VA, and Lefèvre F. (2013) Chemistry of the Atmospheres of Mars, Venus, and Titan. *Comparative Climatology of Terrestrial Planets*, 1, 231-275.

Langmuir D. (1997) Aqueous environmental geochemistry. Upper Saddle River, NJ: Prentice Hall.

Lederberg J. (1965) Signs of life. *Nature*, 207, 9-13.

Léger A. (2000) Strategies for remote detection of life—DARWIN-IRSI and TPF missions—. *Advances in Space Research*, 25, 2209-2223.

Lewis GN, and Randall M. (1923) Thermodynamics and the free energy of chemical substances.

Lewis J. (2012) Physics and chemistry of the solar system. Academic Press.

Line MR, and Yung YL. (2013) A Systematic Retrieval Analysis of Secondary Eclipse Spectra. III. Diagnosing Chemical Disequilibrium in Planetary Atmospheres. *The Astrophysical Journal*, 779, 3.

Lippincott ER, Eck RV, Dayhoff MO, and Sagan C. (1967) Thermodynamic equilibria in planetary atmospheres. *The Astrophysical Journal*, 147, 753.

Lodders K, and Fegley B. (1998) The planetary scientist's companion. Oxford University Press, New York.

Lorenz RD, Kirk RL, Hayes AG, Anderson YZ, Lunine JI, Tokano T, Turtle EP, Malaska MJ, Soderblom JM, and Lucas A. (2014) A radar map of Titan Seas: Tidal dissipation and ocean mixing through the throat of Kraken. *Icarus*, 237, 9-15.

Lovelock JE. (1965) A physical basis for life detection experiments. *Nature*, 207, 568-70.

Lovelock JE. (1975) Thermodynamics and the recognition of alien biospheres. *Proceedings of the Royal Society of London B: Biological Sciences*, 189, 167-181.

Lovelock JE, and Margulis L. (1974) Atmospheric homeostasis by and for the biosphere: the Gaia hypothesis. *Tellus*, 26, 2-10.

Lwin Y. (2000) Chemical equilibrium by Gibbs energy minimization on spreadsheets. *International Journal of Engineering Education*, 16, 335-339.

Mahaffy PR, Webster CR, Atreya SK, Franz H, Wong M, Conrad PG, Harpold D, Jones JJ, Leshin LA, and Manning H. (2013) Abundance and isotopic composition of gases in the martian atmosphere from the Curiosity rover. *Science*, 341, 263-266.

Marion GM. (2002) A molal-based model for strong acid chemistry at low temperatures (< 200 to 298 K). *Geochimica et Cosmochimica Acta*, 66, 2499-2516.

Marion GM, and Kargel JS. (2007) Cold aqueous planetary geochemistry with FREZCHEM: from modeling to the search for life at the limits. Springer Science & Business Media.

Melosh HJ. (2011) Planetary surface processes. Cambridge University Press.

Misra A, Meadows V, Claire M, and Crisp D. (2014) Using dimers to measure biosignatures and atmospheric pressure for terrestrial exoplanets. *Astrobiology*, 14, 67-86.

Moses JI, Visscher C, Fortney JJ, Showman AP, Lewis NK, Griffith CA, Klippenstein SJ, Shabram M, Friedson AJ, and Marley MS. (2011) Disequilibrium carbon, oxygen, and nitrogen chemistry in the atmospheres of HD 189733b and HD 209458b. *The Astrophysical Journal*, 737, 15.

Nair H, Allen M, Anbar AD, Yung YL, and Clancy RT. (1994) A photochemical model of the Martian atmosphere. *Icarus*, 111, 124-150.
Nichita DV, Gomez S, and Luna E. (2002) Multiphase equilibria calculation by direct minimization of Gibbs free energy with a global optimization method. *Computers & chemical engineering*, 26, 1703-1724.
Perry RH, Green DW, and Knovel (Firm). (2008) Perry's chemical engineers' handbook. McGraw-Hill, New York.
Pilson ME. (2012) An Introduction to the Chemistry of the Sea. Cambridge University Press.
Pont F, Knutson H, Gilliland R, Moutou C, and Charbonneau D. (2008) Detection of atmospheric haze on an extrasolar planet: the 0.55–1.05 μm transmission spectrum of HD 189733b with the Hubble Space Telescope. *Monthly Notices of the Royal Astronomical Society*, 385, 109-118.
Prausnitz JM, Lichtenthaler RN, and de Azevedo EG. (1999) Molecular thermodynamics of fluid-phase equilibria. Pearson Education.
Prinn R, Huang J, Weiss R, Cunnold D, Fraser P, Simmonds P, McCulloch A, Harth C, Salameh P, and O'Doherty S. (2001) Evidence for substantial variations of atmospheric hydroxyl radicals in the past two decades. *Science*, 292, 1882-1888.
Prinn RG, and Barshay SS. (1977) Carbon monoxide on Jupiter and implications for atmospheric convection. *Science*, 198, 1031-1034.
Rakov VA, and Uman MA. (2007) Lightning: physics and effects. Cambridge University Press.
Rauer H, Gebauer Sv, Paris P, Cabrera J, Godolt M, Grenfell J, Belu A, Selsis F, Hedelt P, and Schreier F. (2011) Potential biosignatures in super-Earth atmospheres: I. Spectral appearance of super-Earths around M dwarfs. *Astronomy and astrophysics*, 529.
Robinson TD, and Catling DC. (2014) Common 0.1 bar tropopause in thick atmospheres set by pressure-dependent infrared transparency. *Nat. Geosci.*, 7, 12-15.
Robinson TD, Meadows VS, and Crisp D. (2010) Detecting oceans on extrasolar planets using the glint effect. *The Astrophysical Journal Letters*, 721, L67.
Rodler F, and López-Morales M. (2014) Feasibility Studies for the Detection of O2 in an Earth-like Exoplanet. *The Astrophysical Journal*, 781, 54.
Sagan C, Thompson WR, Carlson R, Gurnett D, and Hord C. (1993) A search for life on Earth from the Galileo spacecraft. *Nature*, 365, 715-721.
Schwartzman D, and Volk T. (2004) Does life drive disequilibrium in the biosphere. *Scientists Debate Gaia: The Next Century*, 129-135.
Schwieterman EW, Robinson TD, Meadows VS, Misra A, and Domagal-Goldman S. (2015) Detecting and Constraining N2 Abundances in Planetary Atmospheres Using Collisional Pairs. *The Astrophysical Journal*, 810, 57.

Seager S. (2014) The future of spectroscopic life detection on exoplanets. *Proceedings of the National Academy of Sciences*, 111, 12634-12640.

Seager S, and Bains W. (2015) The search for signs of life on exoplanets at the interface of chemistry and planetary science. *Science Advances*, 1, e1500047.

Seager S, Bains W, and Hu R. (2013) A biomass-based model to estimate the plausibility of exoplanet biosignature gases. *The Astrophysical Journal*, 775, 104.

Seager S, and Deming D. (2010) Exoplanet Atmospheres. *Annual Review of Astronomy and Astrophysics*, 48, 631-672.

Sillén LG. (1966) Regulation of O2, N2 and CO2 in the atmosphere; thoughts of a laboratory chemist. *Tellus*, 18, 198-206.

Simoncini E, Virgo N, and Kleidon A. (2013) Quantifying drivers of chemical disequilibrium: theory and application to methane in the Earth's atmosphere. *Earth System Dynamics*, 4, 317-331.

Sleep NH. (2005) Dioxygen over geological time. *Metal ions in biological systems, biogeocehmical cycles of elements Sigel A, Sigel H, Sigel RKO*, 43, 49-73.

Smirnov A, Hausner D, Laffers R, Strongin DR, and Schoonen MA. (2008) Abiotic ammonium formation in the presence of Ni-Fe metals and alloys and its implications for the Hadean nitrogen cycle. *Geochem Trans*, 9.

Snellen I, de Kok R, Le Poole R, Brogi M, and Birkby J. (2013) Finding extraterrestrial life using ground-based high-dispersion spectroscopy. *The Astrophysical Journal*, 764, 182.

Stevenson KB, Harrington J, Nymeyer S, Madhusudhan N, Seager S, Bowman WC, Hardy RA, Deming D, Rauscher E, and Lust NB. (2010) Possible thermochemical disequilibrium in the atmosphere of the exoplanet GJ 436b. *Nature*, 464, 1161-1164.

Summers DP, Basa RC, Khare B, and Rodoni D. (2012) Abiotic nitrogen fixation on terrestrial planets: Reduction of NO to ammonia by FeS. *Astrobiology*, 12, 107-114.

Ulanowicz RE, and Hannon B. (1987) Life and the production of entropy. *Proceedings of the Royal society of London. Series B. Biological sciences*, 232, 181-192.

Venot O, Hébrard E, Agúndez M, Dobrijevic M, Selsis F, Hersant F, Iro N, and Bounaceur R. (2013) The nitrogen chemistry in hot Jupiters atmosphere. In: *The Early Evolution of the Atmospheres of Terrestrial Planets*, Springer, p 67-83.

Vidal-Madjar A, Des Etangs AL, Désert J-M, Ballester G, Ferlet R, Hébrard G, and Mayor M. (2003) An extended upper atmosphere around the extrasolar planet HD209458b. *Nature*, 422, 143-146.

Walas SM. (1985) Phase equilibria in chemical engineering. Butterworth, Boston.

Walther JV. (2009) Essentials of geochemistry. Jones & Bartlett Publishers.

Wedepohl KH. (1995) The composition of the continental crust. *Geochimica et cosmochimica Acta*, 59, 1217-1232.

White WB, Johnson SM, and Dantzig GB. (1958) Chemical equilibrium in complex mixtures. *The Journal of Chemical Physics*, 28, 751-755.

Younglove B, and Ely JF. (1987) Thermophysical properties of fluids. II. Methane, ethane, propane, isobutane, and normal butane. *Journal of Physical and Chemical Reference Data*, 16, 577-798.

Yung YL, and DeMore WB. (1999) Photochemistry of planetary atmospheres. Oxford University Press, New York.

Zahnle K, Haberle RM, Catling DC, and Kasting JF. (2008) Photochemical instability of the ancient Martian atmosphere. *Journal of Geophysical Research: Planets (1991–2012)*, 113.

Zhang X, Liang MC, Mills FP, Belyaev DA, and Yung YL. (2012) Sulfur chemistry in the middle atmosphere of Venus. *Icarus*, 217, 714-739.

Zugger ME, Kasting JF, Williams DM, Kane TJ, and Philbrick CR. (2010) Light scattering from exoplanet oceans and atmospheres. *The Astrophysical Journal*, 723, 1168.

**Tables**

Table 1: Equilibrium calculation for Venus' atmosphere (T=735.3 K, P=92.1 bar). The second column gives the observed surface mixing ratios of all species in Venus' atmosphere and the third column gives the equilibrium abundances of each species as determined by our own Gibbs free energy minimization Matlab code. The fourth column is an independent validation of the equilibrium abundances calculated using the commercial software package, Aspen Plus. The fifth column gives the change in abundance for each species according to our Gibbs energy minimization (column three minus column two). Bolded rows highlight the species where abundances change during the reaction to equilibrium. The available Gibbs energy from our own code is $\Phi = 0.0596$ J/mol.

| Species | Initial mixing ratio | Final abundance (*fmincon*) | Final abundance (*Aspen*) | Final – initial abundance (*fmincon*) |
|---|---|---|---|---|
| **$CO_2$** | **0.965** | **0.965004** | **0.9650041** | **0.0000039** |
| $N_2$* | 0.034715 | 0.034715 | 0.034715 | -6.21E-13 |
| **$SO_2$** | **0.00015** | **0.000148** | **0.000147949** | **-0.00000197** |
| **$H_2O$** | **0.00003** | **3.00329E-05** | **3.00295E-05** | **3.29E-08** |
| Ar | 0.000061 | 0.000061 | 0.000061 | 0 |
| **CO** | **0.000017** | **1.07569E-05** | **1.04986E-05** | **-0.00000624** |
| He | 0.000009 | 0.000009 | 0.000009 | 0 |
| Ne | 0.000007 | 0.000007 | 0.000007 | 0 |
| **OCS** | **0.00001** | **1.23452E-05** | **1.24294E-05** | **0.00000235** |
| **$H_2S$** | **0.00000007** | **4.00713E-08** | **4.13474E-08** | **-2.99E-08** |
| HCl | 0.0000005 | 0.0000005 | 0.0000005 | 0 |
| Kr | 0.000000025 | 0.000000025 | 0.000000025 | 0 |
| **S** | **0.00000035** | **1.34652E-10** | **2.1696E-17** | **-0.00000035** |
| HF | 4.5E-09 | 4.5E-09 | 4.5E-09 | 0 |
| Xe | 0.00000002 | 0.00000002 | 0.00000002 | 0 |
| **$H_2$** | **0.000000003** | **7.5802E-11** | **2.13576E-09** | **-2.92E-09** |
| **$NH_3$** | **1E-14** | **1.25215E-12** | **1.1679E-14** | **1.24E-12** |

*N2 was slightly modified from textbook value to ensure mixing ratios summed to 1.

Table 2: Equilibrium calculation for Mars' atmosphere (T=214K, P= 0.006 bar). Columns are the same as in table 1. The initial mixing ratios are surface abundances. The available Gibbs energy from our own code is $\Phi = 136.3$ J/mol.

| Species | Initial mixing ratio | Final abundance (*fmincon*) | Final abundance (*Aspen*) | Final – initial abundance (*fmincon*) |
|---|---|---|---|---|
| $CO_2$ | **0.9597** | **0.960257** | **0.960257** | **0.000557** |
| $N_2$ | 0.0189 | 0.0189 | 0.0189 | 5E-10 |
| Ar* | 0.019165 | 0.019165 | 0.0191646 | 0 |
| $O_2$ | **0.00146** | **0.001175** | **0.00117462** | **-0.00028538** |
| CO | **0.000557** | **5.51991E-17** | **0** | **-0.000557** |
| $H_2O$ | **0.0002** | **0.000215** | **0.00021504** | **0.000015** |
| NO | **0.000000001** | **2.36011E-16** | **1.96E-24** | **-0.000000001** |
| Ne | 0.0000025 | 0.0000025 | 0.0000025 | 0 |
| Kr | 0.0000003 | 0.0000003 | 0.0000003 | 0 |
| Xe | 0.00000008 | 0.00000008 | 0.00000008 | 0 |
| $O_3$ | **0.0000004** | **8.702E-17** | **0** | **-0.0000004** |
| $NO_2$ | 1E-30 | 8.84675E-16 | 6.19E-17 | 8.85E-16 |
| $H_2$ | **0.000015** | **6.01993E-17** | **0** | **-0.000015** |
| $H_2O_2$ | **0.00000004** | **1.19914E-16** | **0** | **-0.00000004** |

*Ar was modified slightly from textbook value to ensure mixing ratios summed to 1.

Table 3: Equilibrium calculation for Jupiter's atmosphere. Columns are the same as in table 1. The initial mixing ratios are abundances at the 1 bar level (T=165K). The available Gibbs energy from our own code is $\Phi = 0.00103$ J/mol.

| Species | Initial mixing ratio | Final abundance (*fmincon*) | Final abundance (*Aspen*) | Final – initial abundance (*fmincon*) |
|---|---|---|---|---|
| $H_2$ | **0.862** | **0.86199998** | **0.862** | **-1.56E-08** |
| He* | 0.136024 | 0.136024 | 0.136024 | 0 |
| $CH_4$ | **0.00181** | **0.001810005** | **0.00181001** | **5.2E-09** |
| $NH_3$ | **0.00013** | **0.000130004** | **0.000130004** | **3.6E-09** |
| Ne | 0.0000199 | 0.0000199 | 0.0000199 | 0 |
| Ar | 0.0000157 | 0.0000157 | 0.0000157 | 0 |
| $H_2O$ | **0.000000001** | **2.6E-09** | **2.6E-09** | **1.6E-09** |
| CO | **1.6E-09** | **1.13471E-19** | **0** | **-1.6E-09** |
| HCN | **3.6E-09** | **9.49988E-20** | **0** | **-3.6E-09** |

*He was modified from textbook value to ensure mixing ratios summed to 1.

Table 4: Equilibrium calculation for Titan's atmosphere (T=93.65K, P=1.46 bar). Columns are the same as in table 1. The initial mixing ratios are surface abundances. The available Gibbs energy from our own code is $\Phi = 1.21$ J/mol. NA indicates that these species were not included in the Aspen Plus calculation.

| Species | Initial mixing ratio | Final abundance (*fmincon*) | Final abundance (*Aspen*) | Final – initial abundance (*fmincon*) |
|---|---|---|---|---|
| $N_2$* | 0.94179679 | 0.94179679 | 0.9417968 | 0 |
| $CH_4$ | **0.05712** | **0.057144** | **0.057144** | **0.000024** |
| $H_2$ | **0.00099** | **0.000974** | **0.000974** | **-0.000016** |
| CO | **0.000047** | **4.7E-05** | 0.000047 | -3.46E-19 |
| Ar | 0.00003421 | 0.00003421 | 0.0000342 | 0 |
| $C_2H_6$ | **0.00001** | **3.97606E-19** | **0** | **-0.00001** |
| $C_2H_2$ | **0.000002** | **6.1423E-20** | **0** | **-0.000002** |
| HCN | 1E-20 | 1.15092E-19 | 0 | 1.05E-19 |
| $C_3H_8$ | 1E-20 | 1.9732E-19 | 0 | 1.87E-19 |
| $C_2H_4$ | 1E-20 | 1.1886E-19 | 0 | 1.09E-19 |
| $CH_3C_2H$ | 1E-20 | 5.65276E-20 | NA | 4.65E-20 |
| $C_2N_2$ | 1E-20 | 4.97628E-20 | 0 | 3.98E-20 |
| $C_3HN$ | 1E-20 | 3.86518E-20 | NA | 2.87E-20 |
| $CH_3CN$ | 1E-20 | 1.11588E-19 | NA | 1.02E-19 |
| $C_4H_8\_I$ | 1E-20 | 8.17504E-20 | NA | 7.18E-20 |
| $C_4H_{10}\_I$ | 1E-20 | 1.38594E-19 | NA | 1.29E-19 |
| $C_6H_6$ | 1E-20 | 2.69708E-20 | 0 | 1.7E-20 |
| $NO_2$ | 1E-20 | 1.18993E-19 | 0 | 1.09E-19 |
| NO | 1E-20 | 1.36611E-19 | 0 | 1.27E-19 |

*N2 was modified from textbook value to ensure mixing ratios summed to 1.

Table 5a: Equilibrium calculation for Uranus' atmosphere. Columns are the same as in table 1. The initial mixing ratios are abundances at the 1 bar level (T=75K). The available Gibbs energy from our own code is $\Phi = 0$ J/mol.

| Species | Initial mixing ratio | Final abundance (*fmincon*) | Final abundance (*Aspen*) | Final − initial abundance (*fmincon*) |
|---|---|---|---|---|
| $H_2$ | 0.825 | 0.825 | 0.825 | 0 |
| He* | 0.1519987 | 0.1519987 | 0.1519987 | 0 |
| $CH_4$ | 0.023 | 0.023 | 0.023 | 0 |
| $NH_3$ | 1E-15 | 1E-15 | 0 | 0 |
| $H_2S$ | 0.0000008 | 0.0000008 | 0.0000008 | 0 |
| CO | 0.0000005 | 0.0000005 | 0.0000005 | 0 |

*He was modified from textbook value to ensure mixing ratios summed to 1.

Table 5b: Equilibrium calculation for Uranus' atmosphere with all stratospheric trace species included. Columns are the same as in table 1. The calculation is performed at P=1 bar and T=75K despite the inclusion of stratospheric species to give an upper bound on the free energy at the 1 bar level. The available Gibbs energy from our own code (with all traced species included) is $\Phi = 0.0971$ J/mol.

| Species | Initial mixing ratio | Final abundance (*fmincon*) | Final abundance (*Aspen*) | Final − initial abundance (*fmincon*) |
|---|---|---|---|---|
| **$H_2$** | **0.825** | **0.82499846** | **0.8249985** | **-0.00000154** |
| He* | 0.15199857 | 0.15199857 | 0.1519986 | 0 |
| **$CH_4$** | **0.023** | **0.02300054** | **0.0230005** | **0.00000054** |
| $NH_3$ | 0.0000001 | 0.0000001 | 0.0000001 | 0 |
| $H_2S$ | 0.0000008 | 0.0000008 | 0.0000008 | 0 |
| **CO** | **0.0000005** | **1.0905E-19** | **0** | **-0.0000005** |
| **$H_2O$** | **0.000000006** | **5.06E-07** | **0.000000506** | **0.0000005** |
| **$C_2H_6$** | **0.00000001** | **3.4293E-19** | **0** | **-0.00000001** |
| **$C_2H_2$** | **0.00000001** | **5.77147E-20** | **0** | **-0.00000001** |

*He was modified from the textbook value to ensure mixing ratios summed to 1.

Table 6: Purely gas phase equilibrium calculation for Earth's atmosphere (ocean not included). Columns are the same as in table 1. The initial mixing ratios are surface abundances (T=288.15K, P= 1.013 bar). The available Gibbs energy for Earth (atmosphere only) from our code is $\Phi = 1.51$ J/mol.

| Species | Initial mixing ratio | Final abundance (*fmincon*) | Final abundance (*Aspen*) | Final – initial abundance (*fmincon*) |
|---|---|---|---|---|
| $N_2$ | **0.773095598** | **0.773095914** | **0.7730921** | **3.16826E-07** |
| $O_2$ | **0.2073826** | **0.2073791** | **0.2073829** | **-3.46776E-06** |
| $H_2O$ | **0.00990082** | **0.00990473** | **0.00990473** | **3.91082E-06** |
| Ar | 0.009247366 | 0.009247366 | 0.00924737 | 0 |
| $CO_2$ | **0.000346529** | **0.000348336** | 0.000348336 | 1.8069E-06 |
| Ne | 1.79997E-05 | 1.79997E-05 | 0.000018 | 0 |
| He | 5.18803E-06 | 5.18803E-06 | 0.00000519 | -2.5411E-21 |
| $CH_4$ | **1.68314E-06** | **2.5343E-20** | **1.13E-48** | **-1.68314E-06** |
| Kr | 1.12869E-06 | 1.12869E-06 | 0.00000113 | 0 |
| $H_2$ | **5.4455E-07** | **1.0381E-19** | **4.08E-32** | **-5.44545E-07** |
| $N_2O$ | **3.16826E-07** | **3.3401E-19** | **3.28E-20** | **-3.16826E-07** |
| CO | **1.2376E-07** | **8.7068E-20** | **2.18E-32** | **-1.2376E-07** |
| Xe | 8.61371E-08 | 8.61371E-08 | 8.61E-08 | 0 |
| $O_3$ | **4.95041E-08** | **1.64391E-19** | **1.97E-30** | **-4.95041E-08** |
| HCl | 9.90082E-10 | 9.90082E-10 | 9.9E-10 | -6.20385E-25 |

*Taken US standard atmosphere (dry) and added 1% water vapor, then renormalized everything to ensure mixing ratios add to 1.

Table 7: Multiphase equilibrium calculation for Earth's atmosphere-ocean system (T=288.15K, P= 1.013 bar). Columns are the same as in table 1. The initial mixing ratios are surface abundances. Aqueous species are italicized. The available Gibbs energy for Earth's atmosphere-ocean system from our code is Φ = 2326 J/mol. NA indicates that these species were not included in the Aspen Plus calculation. Note the large changes in nitrate, $H^+$ ions and oxygen.

| Species | Initial moles | Final abundance (*fmincon*) | Final abundance (*Aspen*) | Final – initial abundance (*fmincon*) |
|---|---|---|---|---|
| $H_2O_{(l)}$ | **436.7881549** | **436.7217842** | **436.709** | **-0.066370669** |
| $O_2$ | **0.207382567** | **0.008094666** | **1.50756E-05** | **-0.199287902** |
| $N_2$ | **0.773095598** | **0.693382141** | **0.69014709** | **-0.079713457** |
| *$NO_3(-)$* | ***0.00023499*** | ***0.159662537*** | ***0.166132*** | ***0.159427547*** |
| *$H(+)$* | ***5.10711E-08*** | ***0.141936633*** | ***0.1484065*** | ***0.141936582*** |
| $H_2O_{(g)}$ | **0.00990082** | **0.01229017** | **0.0119409** | **0.00238935** |
| Ar | 0.009247366 | 0.009247366 | 0.009247366 | 0 |
| $CO_{2(g)}$ | **0.000346529** | **0.010835944** | **0.00943268** | **0.010489415** |
| Ne | 1.79997E-05 | 1.79997E-05 | 1.79997E-05 | 0 |
| He | 5.18803E-06 | 5.18803E-06 | 5.18803E-06 | 0 |
| $CH_4$ | **1.68314E-06** | **2.32128E-13** | **0** | **-1.68314E-06** |
| Kr | 1.12869E-06 | 1.12869E-06 | 1.1287E-06 | 0 |
| $H_2$ | **5.44545E-07** | **1.15773E-12** | **0** | **-5.44544E-07** |
| $N_2O$ | **3.16826E-07** | **1.66811E-13** | **NA** | **-3.16826E-07** |
| CO | **1.2376E-07** | **1.001E-12** | **0** | **-1.23759E-07** |
| Xe | 8.61371E-08 | 8.61371E-08 | 8.61372E-08 | 0 |
| $O_3$ | **4.95041E-08** | **1.43837E-12** | **0** | **-4.95027E-08** |
| HCl | 9.90082E-10 | 2.83979E-10 | 0 | -7.06103E-10 |
| *$Na(+)$* | *3.672916562* | *3.672916562* | *3.672917* | *0* |
| *$K(+)$* | *0.079974781* | *0.079974781* | *0.0799747* | *0* |
| *$Mg(+2)$* | *0.413816618* | *0.413816618* | *0.4138166* | *0* |
| *$Ca(+2)$* | *0.08052309* | *0.08052309* | *0.080523* | *0* |
| *$Sr(+2)$* | *0.000709668* | *0.000709668* | *0.000709669* | *0* |
| *$Cl(-)$* | ***4.275870063*** | ***4.275870063*** | ***4.27587*** | ***7.061E-10*** |
| *$SO_4(-2)$* | *0.221125177* | *0.221125177* | *0.2211252* | *0* |
| *$HCO_3(-)$* | ***0.013911382*** | ***9.08865E-08*** | ***5.59323E-07*** | ***-0.013911291*** |
| *$Br(-)$* | *0.00661104* | *0.00661104* | *0.00661104* | *0* |
| *$B(OH)_3$* | *0.003258522* | *0.003258522* | *NA* | *0* |
| *$F(-)$* | *0.000532643* | *0.000532643* | *0.000532643* | *0* |
| *$CO2_{(aq)}$* | ***7.598E-05*** | ***0.005262085*** | ***0.00666488*** | ***0.005186105*** |
| *$CO3(-2)$* | ***0.001762422*** | ***7.76602E-14*** | ***0*** | ***-0.001762422*** |

| | | | |
|---|---|---|---|
| *OH(-)* | *5.48309E-05* | *7.33759E-12* | *4.1902E-12* | *-5.48309E-05* |

Table 8: Comparison of the available Gibbs free energy, $\Phi$, in Solar System atmospheres (defined in equation (7)). The second column gives $\Phi$ for each atmosphere as determined by our Gibbs energy minimization calculations. The third column is our semi-analytic approximation of the available Gibbs energy calculated from summing the Gibbs energy changes associated with key reactions (see main text and Appendix D). The fourth column is an independent verification of $\Phi$ using the commercial software package, Aspen Plus. The fifth report compares our values to those of Lovelock (1975).

| | Available Gibbs energy, $\Phi$ (J/mol of atmosphere)[†] | Validation, $\Phi$ (J/mol of atmosphere) | | Lovelock (1975) $\Phi$ (J/mol of atmosphere) |
|---|---|---|---|---|
| | | Semi-analytic approximation | Aspen Plus | |
| Venus | 0.059598 | 0.0565586 | 0.060099 | 5 |
| Earth(atm) | 1.51348 | 1.5072 | 1.52564 | Not reported |
| Earth | 2325.76 | 1723.65★ | 2348* | 55000 |
| Mars | 136.3485 | 136.8070 | 136.3506 | 13 |
| Jupiter | 0.001032077 | 0.00103205 | 0.0010228 | <1 |
| Titan | 1.2126495 | 1.212617 | 1.208787 | Not reported |
| Uranus** | 0.0971394 | 0.0983 | 0.09713801 | Not reported |

★The discrepancy between the numerical and semi-analytic results for the Earth is expected because the semi-analytic approximation does not take into account changing water activity. See the main text and table 9 for a more detailed explanation.

*Note that different electrolyte models in Aspen Plus produce slightly different Gibbs energy changes. The available Gibbs energy using the Electrolyte Non-Random Two Liquid (ELECNRTL) model is 2348 J/mol, whereas the PITZER electrolyte returns a Gibbs energy change of 2205 J/mol (see appendix E for a full description of multiphase Aspen Plus calculations).

[†]Calculated at surface pressure and temperature for Venus, Earth, Mars and Titan. Calculated at 1 bar and T = 165 K and T = 75 K for Jupiter and Uranus, respectively.

**Unrealistically includes stratospheric species and gaseous water vapor so this is an upper bound on free energy.

Table 9: Semi-analytic validation of the numerical calculation of the available Gibbs free energy, $\Phi$, in the Earth atmosphere-ocean system.

| Species included in calculation | Available energy, $\Phi$ (J/mol) | |
| --- | --- | --- |
| | Semi-analytic approximation | Numerical calculation (*fmincon*) |
| $N_2$, $O_2$, $H_2O$, $H+$ and $NO_3-$ only. Water activity=1. | 1051 | 1059 |
| Five species above plus carbon-bearing species. Water activity=1. | 1723 | 1716 |
| All species and water activity included. | NA | 2326 |

Table 10: Sensitivity of the numerical calculations of the available Gibbs energy, Φ, in the Earth's atmosphere-ocean system to perturbations in variables that are unobservable or difficult to observe for exoplanets.

|  |  | Available energy, Φ (J/mol) |
|---|---|---|
| Temperature | T= 273.15 K | 1634.78 |
|  | T= 288.15 K | 2325.76 |
|  | T= 298.15 K | 2824.48 |
|  |  |  |
| Pressure | 0.1 bar | 1354.20 |
|  | 1.013 bar | 2325.76 |
|  | 10 bar | 3891.96 |
|  | 1000 bar | 6878.35 |
|  |  |  |
| Ocean pH | 2 | 1983.28 |
|  | 4 | 2314.26 |
|  | 6 | 2325.71 |
|  | 8.187 (Earth) | 2325.76 |
|  | 12 | 2325.65 |
|  |  |  |
| Salinity | 0 mol/kg | 2290.01 |
|  | 1.1 mol/kg | 2325.76 |
|  | 11.1 mol/kg | 2276.40 |
|  |  |  |
| Ocean volume | 0.1 Earth ocean | 413.62 |
|  | 0.5 Earth ocean | 1442.95 |
|  | 1 Earth ocean | 2325.76 |
|  | 2 Earth oceans | 4188.27 |
|  | 10 Earth oceans | 8956.34 |
|  | 50 Earth oceans | 12626.22 |
|  |  |  |

**Figures**

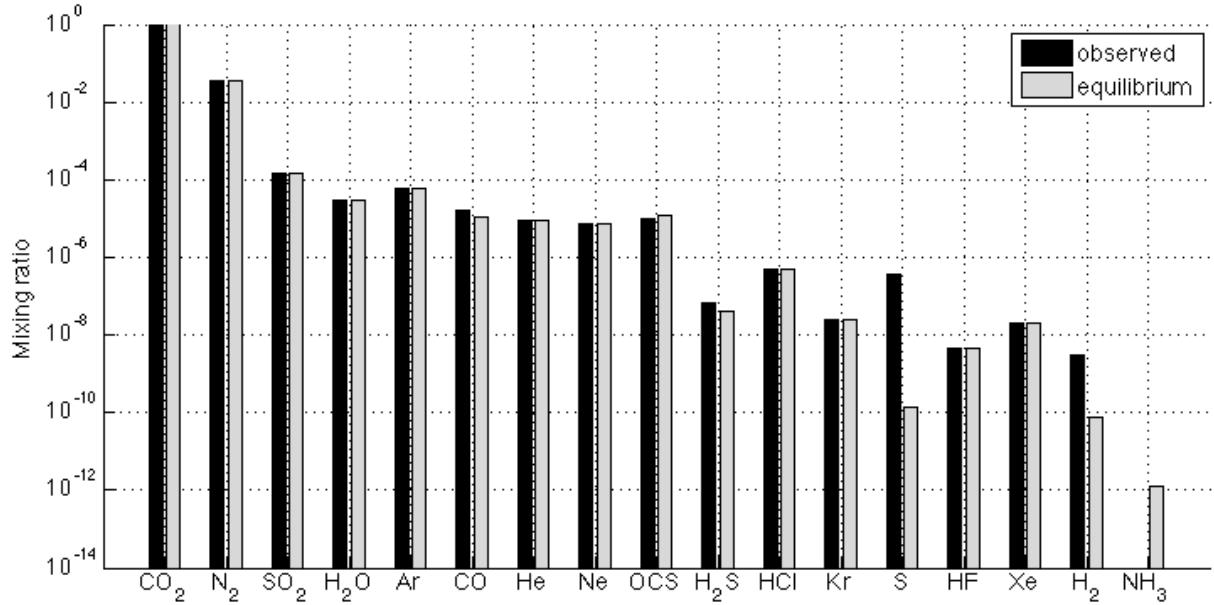

Figure 1: Equilibrium calculation for Venus' atmosphere. The black bars show the observed mixing ratios of all known species in Venus' atmosphere at the surface level ($T$=735.3 K, $P$=92.1 bar). The grey bars show the equilibrium abundances of each of these species as determined by our Gibbs free energy minimization code. The black bars are the column 2 abundances in table 1, and the grey bars are the column 3 abundances in table 1. Notice the loss of S and reduction of CO and $H_2S$ at equilibrium.

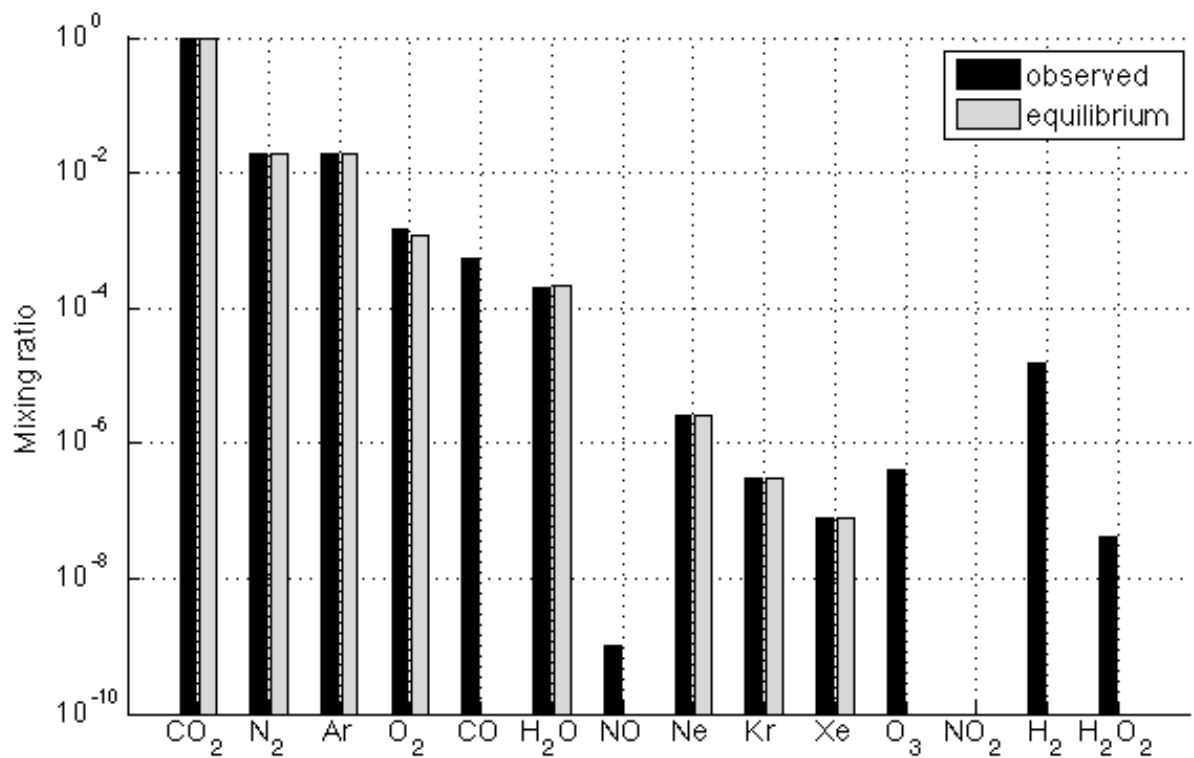

Figure 2: Equilibrium calculation for Mars' atmosphere. The black bars show the observed mixing ratios of all known species in Mars' atmosphere at the surface level (T=214 K, P=0.006 bar). The grey bars show the equilibrium abundances of each of these species as determined by our Gibbs free energy minimization code. The black bars are the column 2 abundances in table 2, and the grey bars are the column 3 abundances in table 2. Notice the loss of CO and reduction of $O_2$ at equilibrium. The compensating increase in $CO_2$ is too small to be visible on this figure.

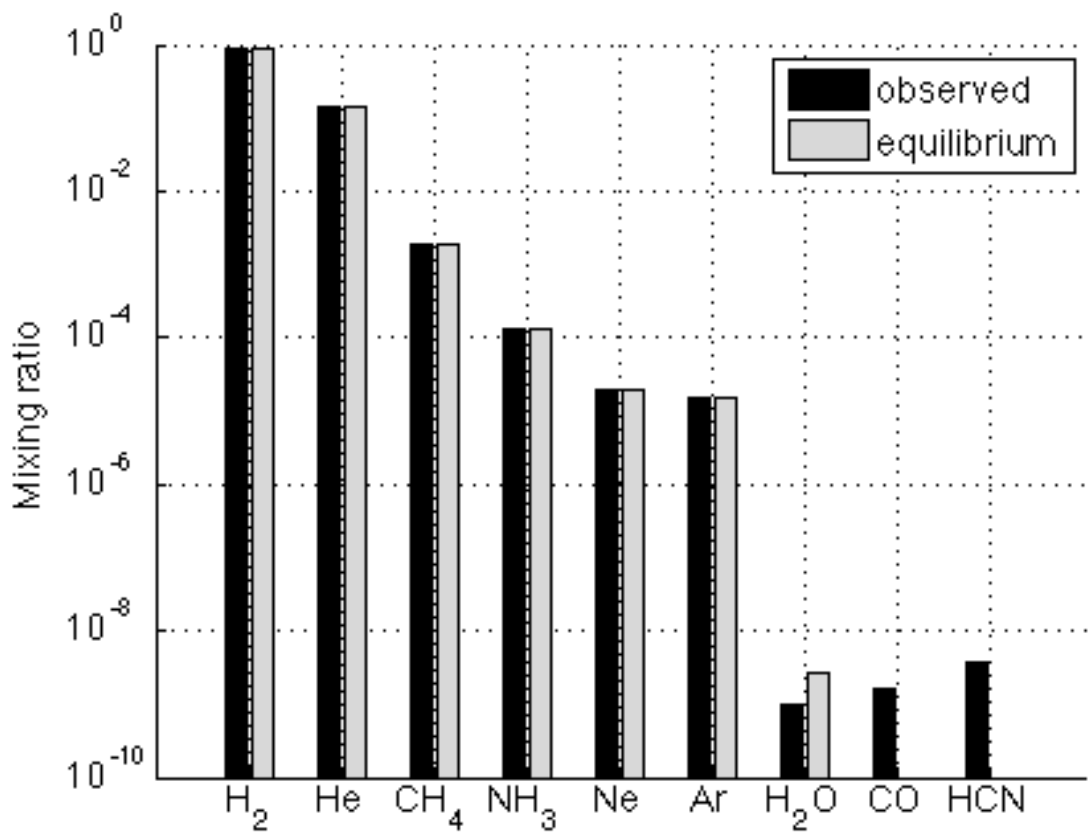

Figure 3: Equilibrium calculation for Jupiter's atmosphere. The black bars show the observed mixing ratios of all known species in Jupiter's atmosphere at the 1 bar level (T=165 K). The grey bars show the equilibrium abundances of each of these species as determined by our Gibbs free energy minimization code. The black bars are the column 2 abundances in table 3, and the grey bars are the column 3 abundances in table 3. Notice the loss of CO and HCN at equilibrium.

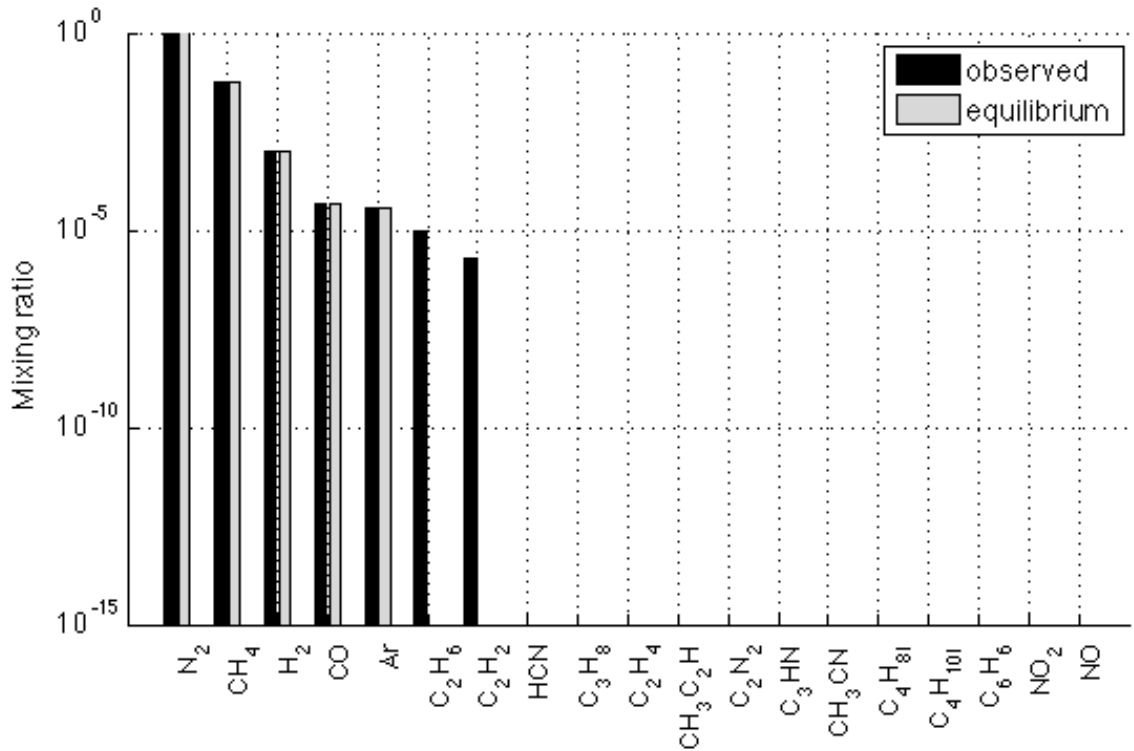

Figure 4: Equilibrium calculation for Titan's atmosphere. The black bars show the observed mixing ratios of all known species in Titan's atmosphere at the surface level (T=93.65K, P=1.46 bar). The grey bars show the equilibrium abundances of each of these species as determined by our Gibbs free energy minimization code. The black bars are the column 2 abundances in table 4, and the grey bars are the column 3 abundances in table 4. Notice the loss of ethane ($C_2H_6$) and acetylene ($C_2H_2$) at equilibrium.

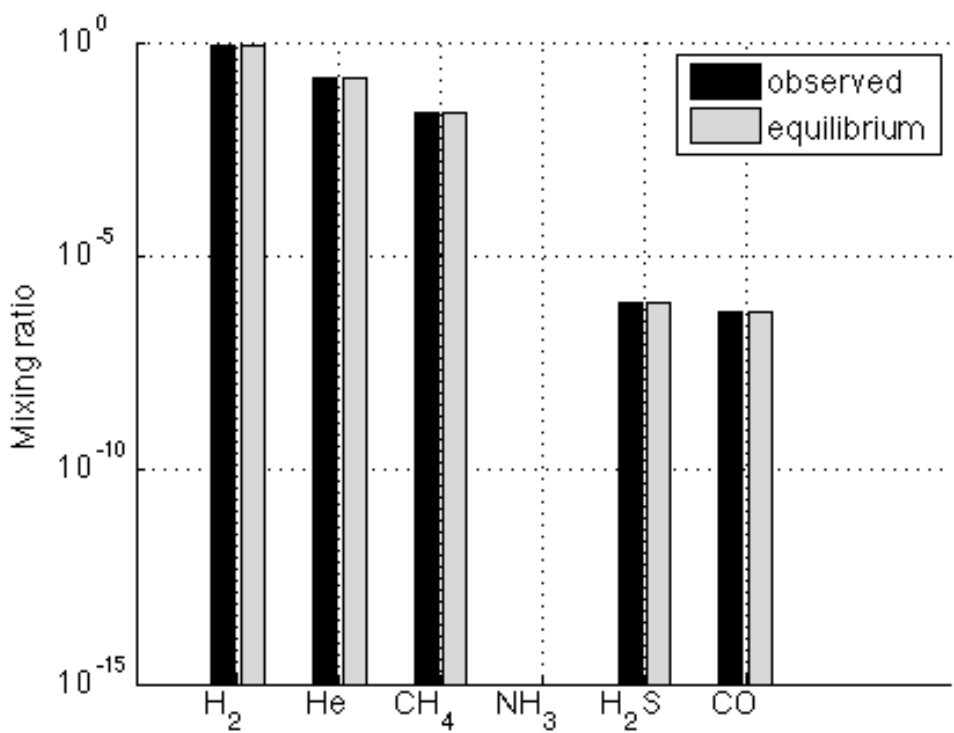

Figure 5: Equilibrium calculation for Uranus' atmosphere. The black bars show the observed mixing ratios of all known species in Uranus' atmosphere at the 1 bar level (T=75K). The grey bars show the equilibrium abundances of each of these species as determined by our Gibbs free energy minimization code. The black bars are the column 2 abundances in table 5a, and the grey bars are the column 3 abundances in table 5a. There is no change in species abundances by reaction to equilibrium.

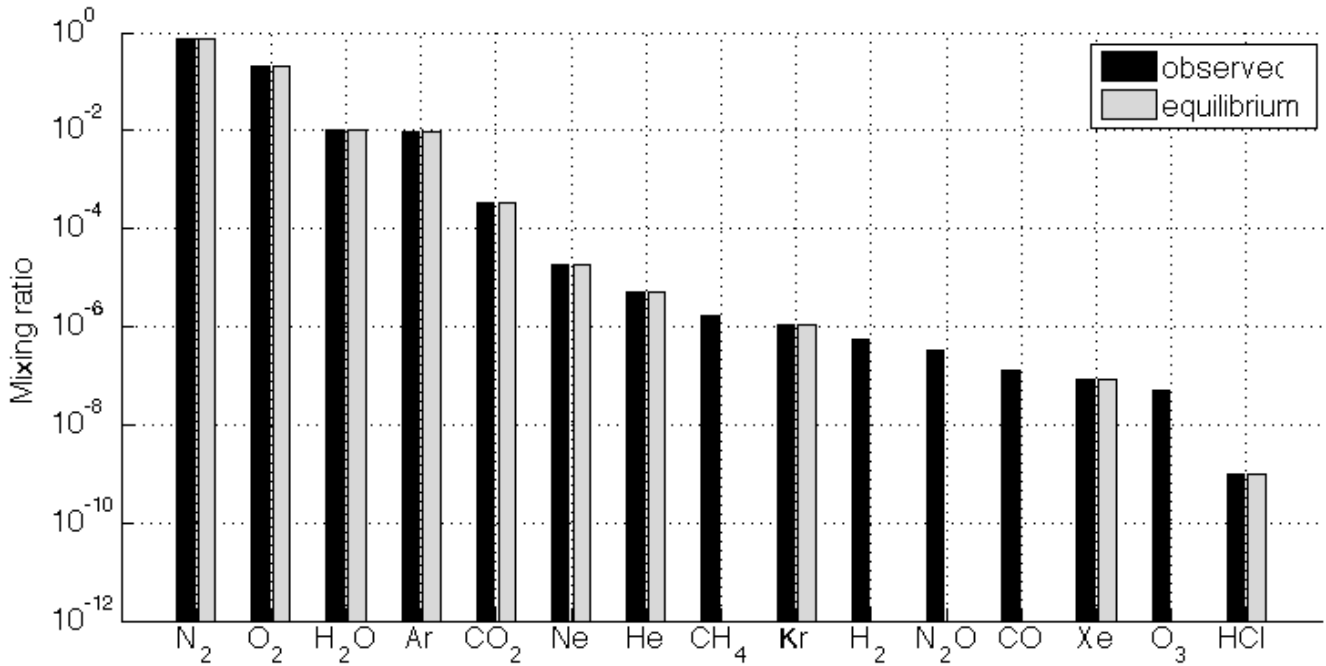

Figure 6: Equilibrium calculation for Earth's atmosphere (not including ocean). The black bars show the observed mixing ratios of all known species in Earth's atmosphere at the surface level (T=288.15 K, P=1.013 bar). The grey bars show the equilibrium abundances of each of these species as determined by our Gibbs free energy minimization code. The black bars are the column 2 abundances in table 6, and the grey bars are the column 3 abundances in table 6. Notice the loss of reduced species ($CH_4$, $H_2$, CO) at equilibrium.

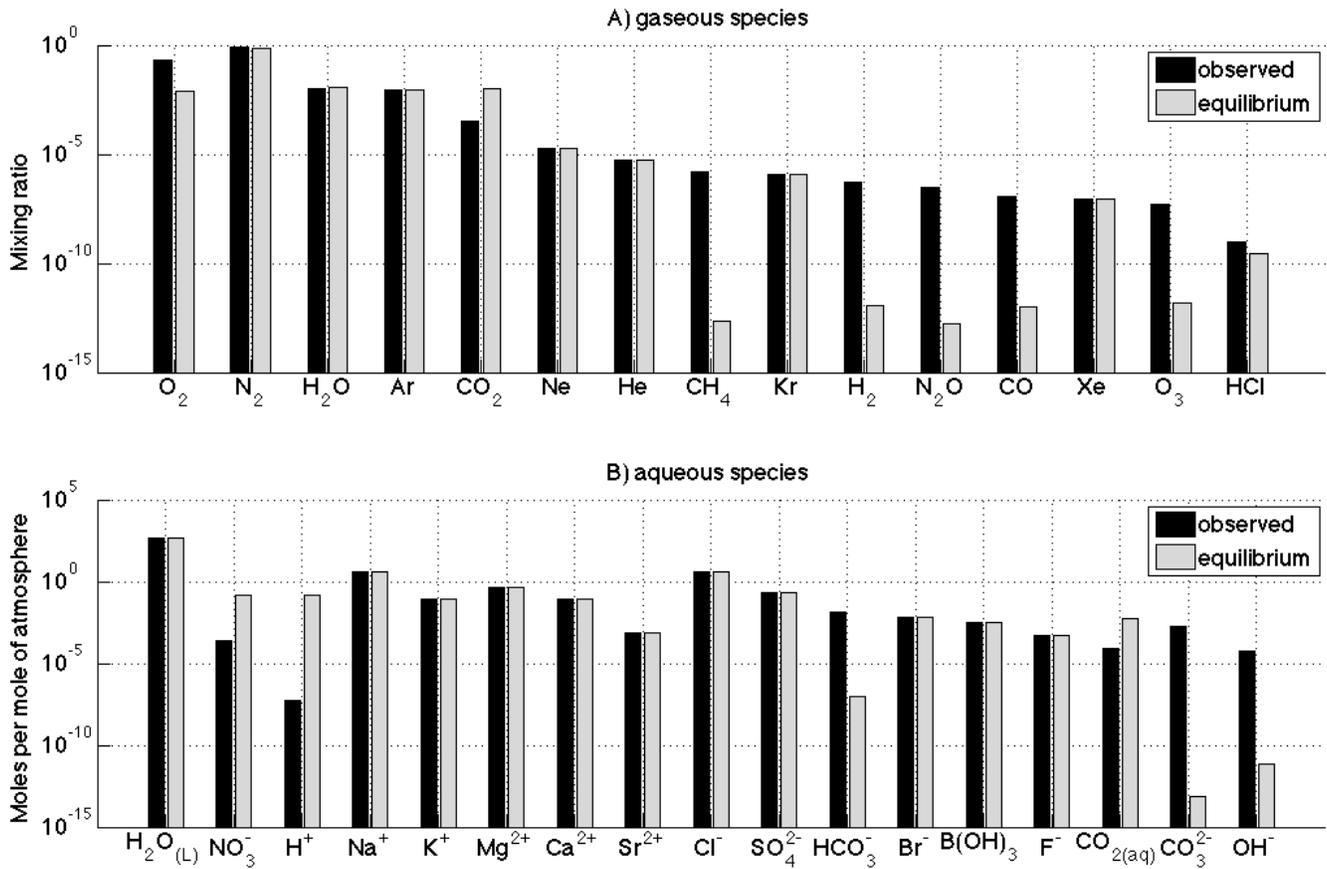

Figure 7: Multiphase equilibrium calculation for Earth's atmosphere-ocean system. The black bars show the observed mixing ratios and abundances of all species in Earth's atmosphere and oceans at the surface level (T=288.15 K, P=1.013 bar). The grey bars show the equilibrium abundances of each of these species as determined by our Gibbs free energy minimization code. The black bars are the column 2 abundances in table 7, and the grey bars are the column 3 abundances in table 7. A) Shows all gas phase species whereas B) shows all aqueous species. Notice that in equilibrium there is a large decrease in $O_2$ since oxygen is converted to nitric acid ($H^+$ and $NO_3^-$ increase) by reaction (21).

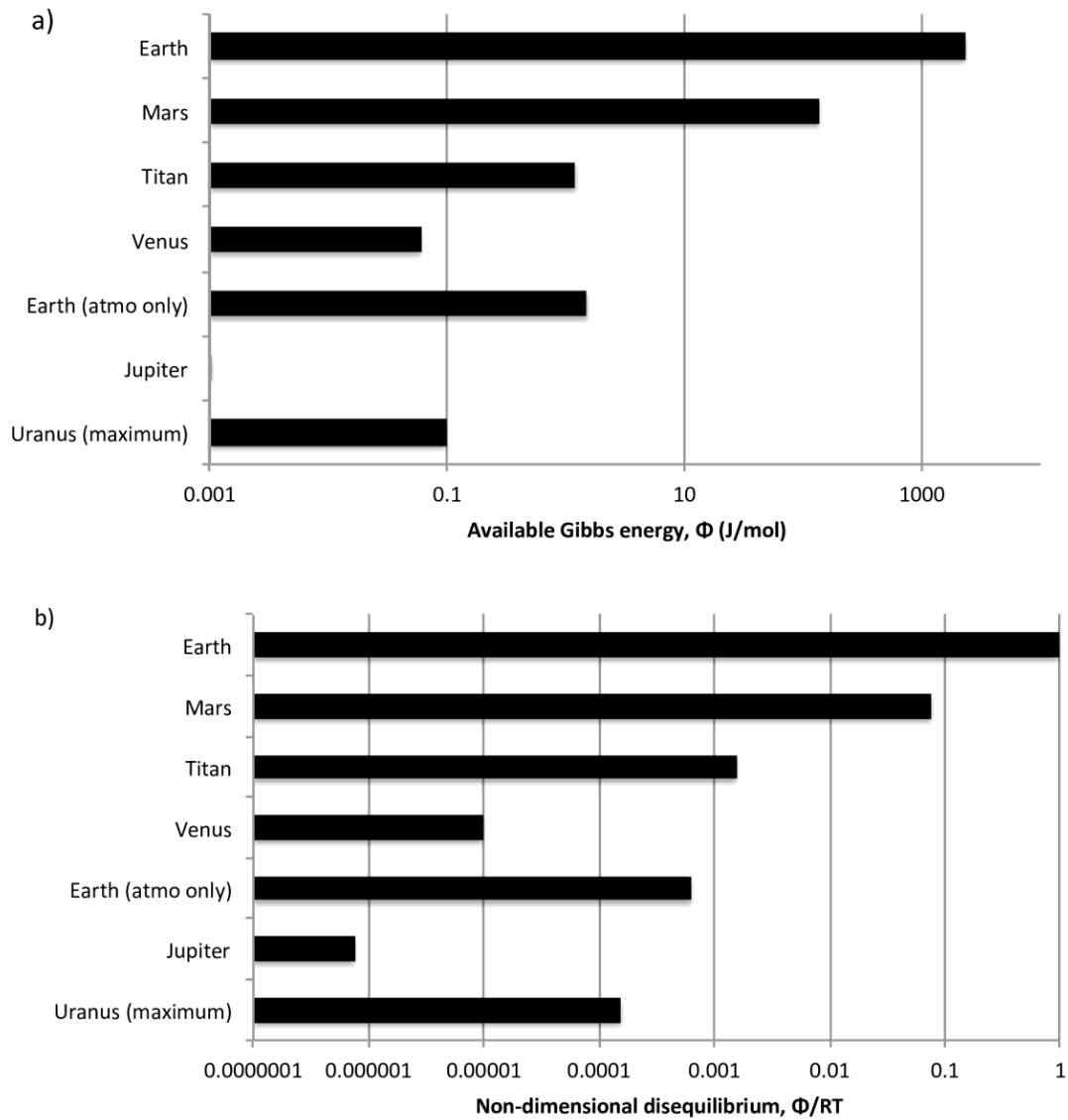

Figure 8: Comparison of the available Gibbs free energy, $\Phi$, in Solar System atmospheres as determined by our Gibbs free energy minimization calculations. The available Gibbs free energies in A) correspond to the second column in table 8. The free energy in the Earth atmosphere-ocean system is more than an order of magnitude greater than any other planetary atmosphere in the Solar System. B) gives the dimensionless free energy for each planet's atmosphere (available Gibbs energy $\Phi$ divided by $RT$). This roughly corrects for the fact that the inner planets receive more free energy from the Sun that is available to drive chemical disequilibrium. Equilibria are calculated at surface pressure and temperature for Venus, Earth, Mars and Titan, and at 1 bar and $T = 165$ K and $T = 75$ K for Jupiter and Uranus, respectively.

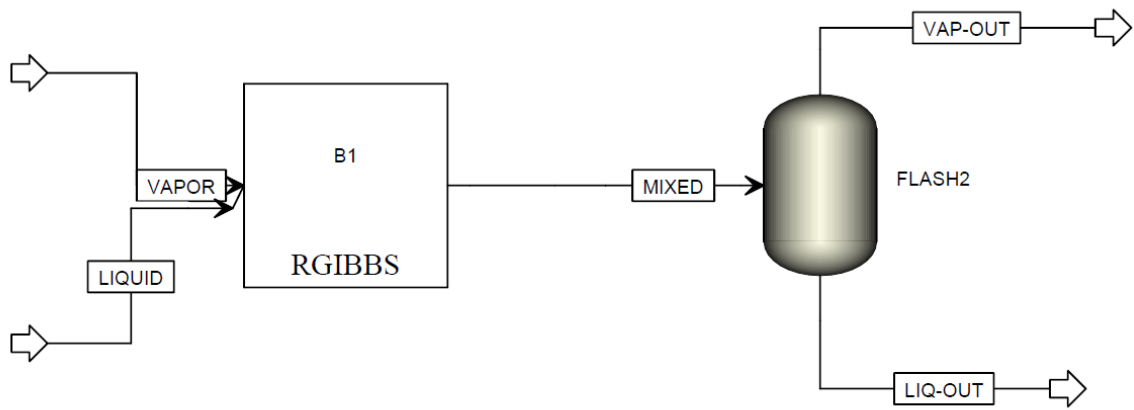

Figure E1: Aspen Plus flowsheet for multiphase calculations.